%% file: main.tex
\input{preamble.tex}
\title{A Persistent Random-Walk Model of Molecular Transport in Neuronal Dendritic Trees}
\author{Lasko Basnarkov$^{1,2}$\\[0.4em]
\small $^1$Faculty of Computer Science and Engineering, Ss. Cyril and Methodius University,\\
\small Skopje, Macedonia\\[0.3em]
\small $^2$Research Center for Computer Science and Information Technologies,\\
\small Macedonian Academy of Sciences and Arts, Skopje, Macedonia\\[0.4em]
\small\href{mailto:lasko.basnarkov@finki.ukim.mk}{lasko.basnarkov@finki.ukim.mk}}
\date{}
\begin{document}
\maketitle
\begin{abstract}
A two-level analytical framework is developed for modeling the stochastic transport of messenger RNA (mRNA) molecules from the soma to synaptic targets along neuronal dendrites. Motivated by experimental observations of intermittent run-and-pause cargo motion, transport along an individual dendritic segment is described by a Persistent Telegraph Process with Pauses. Analytical expressions are derived for segment traversal and return probabilities and the corresponding conditional mean first-passage times under both symmetric and asymmetric directional persistence. A counterintuitive result is that stronger persistence in a given direction does not necessarily lead to faster transport than in the opposite direction. The segment-level quantities are then used to construct a coarse-grained semi-Markov model of mRNA transport on the complete dendritic tree. This model yields target-specific absorption probabilities and conditional mean first-passage times from the soma to individual synaptic targets. Using experimentally motivated transport parameters and empirically reconstructed neuronal morphologies, we identify two distinct asymptotic transport regimes. When retrograde persistence dominates, the conditional mean first-passage times to successfully reached synaptic targets become only weakly dependent on their distance from the soma. In contrast, when anterograde persistence dominates, transport through the dendritic tree effectively resembles motion along a single extended dendritic path, with the conditional mean first-passage time increasing approximately linearly with soma-to-synapse distance. These regimes are theoretically justified when dendritic segments are considerably longer than the distance traveled during a typical run.
\end{abstract}


\section{Introduction}
A central mechanism underlying synaptic plasticity, learning, and memory is local protein synthesis at synaptic regions in response to neuronal activity \cite{richter2009making}. Messenger ribonucleic acid (mRNA) carries genetic information from the nucleus to the cytoplasm, where it can be translated into proteins. Following transcription, many neuronal mRNAs are transported and localized throughout the dendritic arbor, providing local pools of transcripts that can support protein synthesis far from the soma. Although activity-dependent translation can act on mRNA already present near synapses, transport from the soma remains important for establishing and replenishing these localized pools. Classical diffusion may contribute to transport over short distances near the soma, but computer simulations using experimentally measured diffusion coefficients indicate that free diffusion is insufficient to transport mRNAs of sizes comparable to Arc or Ip3r1 to distal dendrites \cite{Fujita2020}. Long-range intracellular transport therefore relies substantially on molecular motors moving cargo along microtubules, primarily kinesins in the anterograde direction and dynein in the retrograde direction \cite{hirokawa2006mrna} (Fig.~1(a)).

Numerous experimental studies of mRNA transport along dendrites have focused on characterizing the dynamics of mRNA cargo motion. It is well established that this motion is stochastic and characterized by intermittent runs and pauses \cite{ahn2023statistical,donlinasp2021differential,bauer2019live}. Observations show that both run and pause durations are random and can vary considerably. During directed runs, the cargo typically moves at approximately constant velocity, with comparable velocity magnitudes in the anterograde and retrograde directions \cite{song2018neuronal}, while the duration and displacement of individual runs remain stochastic \cite{donlinasp2021differential,ahn2023statistical}. During pauses, the cargo remains stationary or exhibits confined fluctuations within a narrow spatial region. Direction reversals and repeated transitions between mobile and immobile states are also observed \cite{bauer2019live,yoon2016glutamate}. Beyond trajectory-level dynamics, experimental contributions have studied the spatial distribution of dendritically localized transcripts, including the dependency of density on the distance from the soma \cite{buxbaum2014single,kim2024spatial}. These experimental observations provide a detailed picture of the local dynamics and spatial organization of dendritic mRNA. However, they do not by themselves determine how such local run--pause dynamics translate into the probability and time required for an individual RNA molecule to reach a particular synaptic region in a branched dendritic tree.

Theoretical studies have addressed stochastic intracellular cargo transport, target search, spatial localization, and first-passage phenomena in neuronal geometries \cite{newby2009directed,newby2010local,williams2016dendritic,jose2018trapping,kreten2026tracking}. In particular, intermittent search models have been used to determine hitting probabilities and conditional first-passage times to synaptic targets on idealized dendritic trees \cite{newby2009directed}, while other approaches have investigated cargo distributions in reconstructed neuronal morphologies \cite{williams2016dendritic} and first-passage transport through branched dendritic structures \cite{jose2018trapping,kreten2026tracking}. However, to the best of our knowledge, a framework combining experimentally characterized run--pause dynamics and directional persistence at the level of individual dendritic segments with target-specific first-passage analysis on reconstructed dendritic morphologies has not yet been developed.

In this manuscript, we develop an experimentally motivated two-level analytical framework that bridges microscopic cargo dynamics and long-range transport in dendritic trees. At the microscopic level, a Persistent Telegraph Process with Pauses (PTPP) \cite{masoliver1989continuous,muller2008motility,bressloff2013stochastic,masoliver2017continuous} is used to model cargo motion along an individual dendritic segment. The model extends the classical persistent random-walk and telegraph-process descriptions \cite{furth1920brownsche,taylor1922diffusion,goldstein1951diffusion,weiss1976two,weiss1994aspects} by introducing four internal states corresponding to anterograde motion, retrograde motion, and the associated paused states. In this way, it captures the principal experimentally observed features of dendritic mRNA transport, including finite-velocity runs, pauses, directional reversals, and directional persistence. The PTPP model allows us to derive segment-level quantities required for transport on the full dendritic tree, including the probabilities of traversing a segment or returning to its point of entry and the corresponding conditional mean first-passage times. In the presence of asymmetric directional persistence, the model predicts an approximately linear scaling of the conditional mean first-passage time with segment length, in contrast to the quadratic distance dependence characteristic of diffusive transport. Moreover, stronger directional persistence does not necessarily imply faster delivery: in one parameter regime, weaker persistence produces a smaller distance-dependent slope of the conditional mean first-passage time.

At the second level, corresponding to the complete dendritic tree, transport is described by a coarse-grained semi-Markov model. The traversal and return probabilities obtained for each dendritic segment determine the transition probabilities of the corresponding embedded Markov chain, while the associated conditional mean first-passage times determine the transition-dependent sojourn times. The states of the semi-Markov process are identified with the soma, dendritic branching points, and synaptic targets, whereas dendritic segments form the links of the corresponding tree network. 
Unlike a continuous-time Markov chain, a semi-Markov description allows the sojourn-time distributions to be non-exponential and, in the present formulation, to depend on both the departure and destination states \cite{smith1955regenerative,levy1956semi, basnarkov2026semi}. Here, the framework is adapted to an absorbing multi-target setting in order to determine target-conditioned first-passage quantities.

The resulting model predicts two distinct asymptotic transport regimes for sufficiently long dendritic segments. When retrograde persistence dominates, the conditional mean first-passage time to successfully reached synaptic targets becomes only weakly dependent on their distance from the soma. In the opposite regime, when anterograde persistence dominates, the transport time grows approximately linearly with soma-to-synapse distance, so that transport through the branched tree effectively resembles transport along a single extended dendritic path. These qualitative regimes are observed across different neuronal morphologies obtained from the public NeuroMorpho.Org repository \cite{Ascoli2007NeuroMorpho}.

The paper is organized as follows. Section 2 develops the theoretical framework, including the PTPP description of transport along individual dendritic segments and the coarse-grained semi-Markov description of transport on the dendritic tree. The symmetric-persistence case and the asymptotic behavior of the two persistence regimes are also analyzed in this section. Section 3 presents numerical results for different neuronal morphologies and discusses their biological interpretation. Finally, Section 4 summarizes the main findings and discusses their implications for mRNA transport in neurons.

\begin{figure*}
\begin{minipage}{0.45\textwidth}
\begin{overpic}[width=\linewidth]{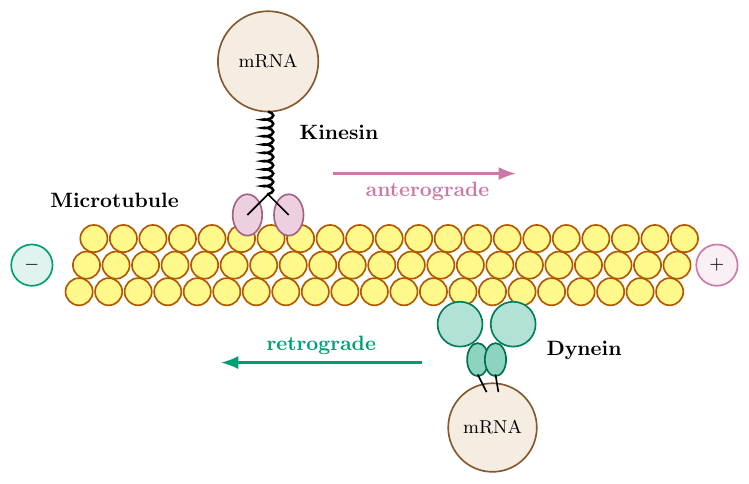}
    \put(10,50){\textbf{(a)}}
\end{overpic}
\end{minipage}
\hfill
\begin{minipage}{0.45\textwidth}
\begin{overpic}[width=\linewidth]{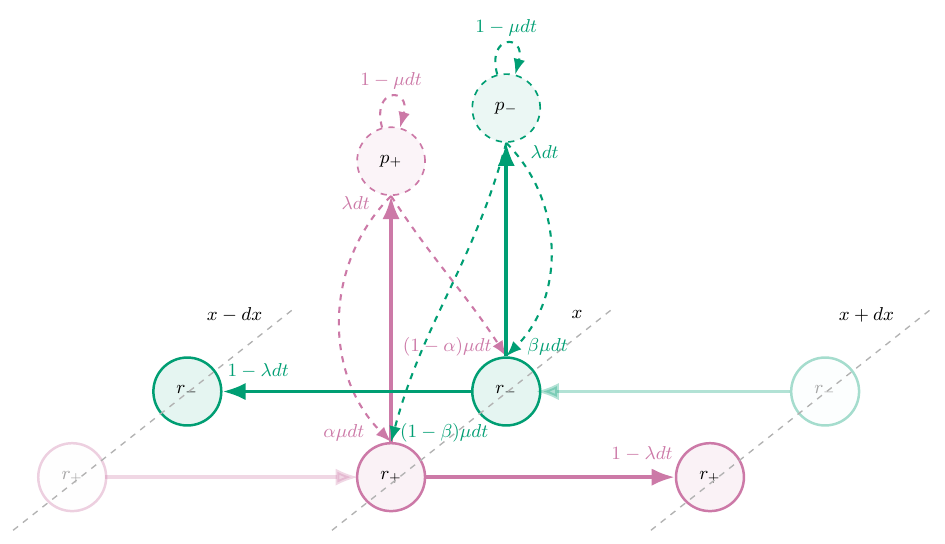}
    \put(10,50){\textbf{(b)}}
\end{overpic}
\end{minipage}

\vspace{0.4cm}

\begin{minipage}{0.65\textwidth}
\begin{overpic}[width=\linewidth]{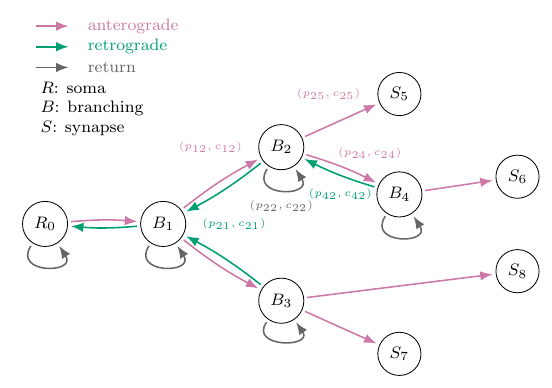}
    \put(-10,70){\textbf{(c)}}
\end{overpic}
\end{minipage} 
\caption{(a) Visual representation of the transport of an mRNA molecule along a microtubule with kinesin and dynein molecular motors. (b) Sketch of the states and transitions, with the corresponding transition rates, in the continuous-time Markov chain model of the Persistent Telegraph Process with Pauses. In this setting, the positive $x$-axis is oriented towards the synapses -- in the anterograde direction. (c) Schematic representation of a small toy neuron model with a few dendritic segments, branching points, and synapses. The transition probabilities and waiting times between nodes are shown only for the node $B_2$. Note that for the loop, the probability and sojourn time are combinations of the respective values along all dendritic segments connected to the node.}
\label{fig:model}
\end{figure*}

\section{Theoretical Model}

\subsection{PTPP Model for mRNA transport along dendrites}

The Persistent Telegraph Process with Pauses on a line represents a natural mathematical model of the experimentally observed random motion of mRNA cargo along dendrites \cite{song2018neuronal}. The PTPP is a continuous-time Markov process in a combined state space with a continuous position coordinate $x$ and discrete internal states $s_m$. For the purpose of the model, the motion is considered to be one-dimensional along the dendritic axis, which is a reasonable approximation since the width of dendrites is much smaller than their length. There are four possible internal states of the process, corresponding to two directions of motion and their associated paused states. The cargo moves with constant velocity $v$ in either direction, and the durations of runs and pauses, for analytical tractability, are exponentially distributed with parameters $\lambda$ and $\mu$, respectively. After each pause, the cargo continues to move in the same direction with a given probability -- $\alpha$ for the anterograde direction and $\beta$ for the retrograde direction. The motion is constrained to the segment $[0, l]$, where $l$ is the dendritic segment length, with absorbing boundaries at $x=0$ and $x=l$. The state of the mRNA cargo at any moment is thus determined by the pair $y = (x, s_m)$, where $s_m \in \mathcal{S}_m = \{r+, r-, p+, p-\}$ denotes the state of motion: running or pausing in the positive or negative direction. The positive direction of motion is that of the $x$-axis, which may coincide with the anterograde or retrograde direction, depending on the specific problem considered.

The dynamics of the state $(x, s_m)$ within the space $[0, l] \times \mathcal{S}_m$ thus follow a continuous-time Markov process, described by transition probabilities over an infinitesimal time interval $dt$, as illustrated in Fig.~\ref{fig:model}(b). Studies of random transport phenomena on a segment are usually focused on two key quantities of interest -- the transition probabilities and the respective MFPTs from a given starting point $x$ to the absorbing boundaries $0$ and $l$. For the purpose of mRNA transport in a cell, one particularly needs the probability $p^{l}(0)$ to reach the end $x=l$ starting from the origin $x=0$ in the positive direction (with $r_+$), and the probability of return to $x=0$, which, under the assumption that the cargo is not destroyed or degraded, is given by the complement $p^{0}(0) = 1 - p^{l}(0)$. The first-passage times of interest are the conditional mean first-passage times (CMFPTs) $c^{l}(0)$ and $c^{0}(0)$, defined as averages over all possible travel times associated with the trajectories (in the combined space $[0, l] \times \mathcal{S}_m$) that end at the farther end $x=l$ or at the origin $x=0$, respectively. For convenience, instead of finding the CMFPTs directly, we calculate the related weighted mean first-passage times (WMFPTs), $w^{l}(0)$ and $w^{0}(0)$. The WMFPT is an average of travel times from a given starting point to a specified absorption site, with weight 0 assigned to trajectories that are absorbed at other targets.

To continue with the analysis, assume that the $x$-axis points in the anterograde direction. Let the PTPP be initialized at time zero from a given state $y = (x, s_m)$ and let $Y_t$ denote a future (random) state at time $t$. Thus, the subscript $t$ denotes the elapsed time from the initialization at $y$. The above-mentioned key quantities can be calculated by using the respective Markov semigroup operator. Its general form is given by 
\begin{equation}
T_t f(y) = \mathbb{E}\left[f(Y_t)|Y_0=y\right] = \mathbb{E}_y\left[f(Y_t)\right],
\end{equation}
where the subscript $y$ in the expectation operator denotes that it is calculated for a process that started at state $y$. The operator $T_t$ determines the expected value at a future time $t$ of a function of the state $f(y)$. A convenient way to obtain the operator is through its associated infinitesimal generator, defined as \cite{ethier2009markov}
\begin{equation}
\mathcal{L}f(y) = \lim_{h \to 0}\frac{T_h f(y)-f(y)}{h}. \label{eq:gen_def}
\end{equation}
The forms of the generator equations for the two quantities of interest, $p$ and $w$, are similar, differing only in the absence or presence of a forcing term. The derivation of the generator uses the transition probabilities and relationships between the states in the continuous-time Markov process, a Taylor expansion of the state function $f$ in the expectation, and the limit $h\to 0$. The details of this standard procedure are provided in the Supplemental Material \cite{SM}. The result is the following set of equations for the generator, with one equation for each of the four states of motion
\begin{eqnarray}
\mathcal{L}f^l_{r+}(x) &=& v\frac{df^l_{r+}(x)}{dx}+\lambda[f^l_{p+}(x)-f^l_{r+}(x)] = \mathcal{F}^l_{r+}(x); \\
\mathcal{L}f^l_{r-}(x) &=& -v\frac{df^l_{r-}(x)}{dx}+\lambda[f^l_{p-}(x)-f^l_{r-}(x)] = \mathcal{F}^l_{r-}(x); \nonumber \\
\mathcal{L}f^l_{p+}(x) &=& \mu[\alpha f^l_{r+}(x)+(1-\alpha)f^l_{r-}(x)-f^l_{p+}(x)] = \mathcal{F}^l_{p+}(x);\nonumber\\ 
\mathcal{L}f^l_{p-}(x) &=& \mu[(1-\beta) f^l_{r+}(x)+\beta f^l_{r-}(x)-f^l_{p-}(x)] = \mathcal{F}^l_{p-}(x)\nonumber, \label{main:eq:L_general}
\end{eqnarray}
where $\mathcal{F}^l(x)$ is the respective forcing function. Note the appearance of a superscript $l$, which denotes that these equations correspond to absorption at $l$. Notice that the last two equations in the system \eqref{main:eq:L_general} are algebraic and express the functions corresponding to the paused states in terms of those for the running states. The systems are solved with the respective boundary conditions $p^l_{r+}(l)=1$ and $p^l_{r-}(0)=0$ for the absorption probability and $w^l_{r+}(l)=0$ and $w^l_{r-}(0)=0$ for the WMFPT. When motion starts from the end of the dendritic segment farther from the soma, the $x$-axis is oriented in the retrograde direction. One can then switch the persistence probabilities $\alpha$ and $\beta$ and use the same relationships. 

The motion of an mRNA cargo along a dendrite is a building block for the semi-Markov model of transport throughout the whole neuron. The respective quantities for each dendrite are needed to construct the transition probabilities and sojourn times in the semi-Markov chain. For this purpose, the motion along a dendrite from branching point $i$ to $j$ is specified by four quantities for each of the two directions of entry into the dendritic segment: anterograde and retrograde. When the cargo enters at the side that is closer to the soma, one needs the following quantities: 1. the probability to traverse the dendrite and exit at the other end -- $p^{l}_{ij}(0)$; 2. the probability of returning -- exiting at the entry point $p^{0}_{ij}(0)$; 3. the CMFPT for traversal in the anterograde direction $c^{l}_{ij}(0)=w^{l}_{ij}(0)/p^{l}_{ij}(0)$; and 4. the CMFPT to return to the entry point $c^{0}_{ij}(0)=w^{0}_{ij}(0)/p^{0}_{ij}(0)$. It should be noted that the starting state of the cargo (on entry into the dendritic segment) is a running state in the positive direction in the coordinate system used for the telegraph process. The quantities corresponding to entry from the opposite end of the dendrite can be conveniently denoted by a bar over the respective variable: $\overline{p}^{l}_{ij}(0), \overline{p}^{0}_{ij}(0), \overline{c}^{l}_{ij}(0), \overline{c}^{0}_{ij}(0)$. For these quantities, the parameters $\alpha$ and $\beta$ are the persistence probabilities for the {\it retrograde} and {\it anterograde} directions, respectively. Obviously, all eight quantities depend on the length of the dendrite between branching points $i$ and $j$, a dependence omitted to simplify the notation.

\subsection{Relationships for the absorption probabilities}

The right-hand sides of the generator equations for the absorption probability are zero, since the probability of reaching the target is neither destroyed nor generated. Thus, the probability $p$ is a harmonic function $\mathcal{L}p=0$ and one has a Dirichlet problem. One can use the last two relationships in \eqref{main:eq:L_general} to express the probabilities of absorption from paused states as functions of those from running states 
\begin{eqnarray}
p^E_{p+}(x) &=& \alpha p^E_{r+}(x)+(1-\alpha)p^E_{r-}(x)\nonumber\\ 
p^E_{p-}(x) &=& (1-\beta) p^E_{r+}(x)+\beta p^E_{r-}(x).\label{main:eq:abs_prob_rest_from_m}
\end{eqnarray}
This results in the following system for the absorption probabilities related to the running states
\begin{eqnarray}
\frac{dp^E_{r+}}{dx} &=& \frac{\lambda}{v}(1-\alpha)[p^E_{r+}(x) - p^E_{r-}(x)]\nonumber\\
\frac{dp^E_{r-}}{dx} &=& \frac{\lambda}{v}(1-\beta)[p^E_{r+}(x) - p^E_{r-}(x)], \label{main:eq:Pi_diff_eq}
\end{eqnarray}
which, after introducing the substitutions $A=(1-\alpha)\lambda/v$ and $B=(1-\beta)\lambda/v$, can be cast in the following matrix form
\begin{equation}
\begin{bmatrix}
\frac{dp^E_{r+}}{dx} \\
\frac{dp^E_{r-}}{dx}
\end{bmatrix} = \begin{bmatrix}
A & -A \\
B & -B 
\end{bmatrix} \cdot \begin{bmatrix}
p^E_{r+}(x) \\
p^E_{r-}(x)
\end{bmatrix}. \label{main:eq:Abs_p_matrix}
\end{equation}
The structure of the matrix equation suggests that it is convenient to introduce the difference $p^E_D(x) = p^E_{r+}(x) - p^E_{r-}(x)$ and the sum of absorption probabilities $p^E_S(x) = p^E_{r+}(x) + p^E_{r-}(x)$. Now the equation for the difference is
\begin{equation}
\frac{dp^E_D}{dx} = (A-B)p^E_D(x),
\end{equation}
with solution
\begin{equation}
p^E_D(x) = C_1 e^{kx},
\end{equation}
where the exponential rate is $k=A-B$. The equation for the sum now becomes 
\begin{equation}
\frac{dp^E_S}{dx} = (A+B)p^E_D(x),
\end{equation}
which, after substituting the difference of probabilities and integrating, gives 
\begin{equation}
p^E_S(x) = \frac{C_1(A+B)}{k} e^{kx} + C_2.
\end{equation}
We now consider absorption at the endpoints $0$ and $l$ as separate cases. The probability of absorption at the right endpoint is obtained from the boundary conditions $p^l_{r+}(l)=1$ (certain absorption at $l$) and $p^l_{r-}(0)=0$ (impossible absorption at $l$).  The resulting expressions are
\begin{equation}
p^l_{r+}(x) = \frac{Ae^{kx}-B}{Ae^{kl}-B}; \hspace{1cm}
p^l_{r-}(x) = \frac{Be^{kx}-B}{Ae^{kl}-B}, 
\end{equation}
which, after substituting $C = A/(Ae^{kl}-B)$ and $D = B/(Ae^{kl}-B)$, take the more convenient form
\begin{equation}
p^l_{r+}(x) = Ce^{kx}-D; \hspace{1cm}
p^l_{r-}(x) = De^{kx}-D. \label{main:eq:abs_prob_l}
\end{equation}
Asymptotically, for $x\to 0$ one has
\begin{equation}
p^l_{r+}(0) = C-D; \hspace{1cm}
p^l_{r-}(0) = 0.
\end{equation}
At the other end of the dendritic segment, $x=l$, the asymptotic behavior for $kl \gg 1$ is given by
\begin{equation}
p^l_{r+}(l) = 1; \hspace{1cm}
p^l_{r-}(l) =\frac{B}{A}=\frac{1-\beta}{1-\alpha}.
\end{equation}
The complementary probabilities of reaching the left endpoint are $p^0_{r+}(x) = 1 - p^l_{r+}(x)$ and $p^0_{r-}(x) = 1 - p^l_{r-}(x)$, or
\begin{equation}
p^0_{r+}(x) = C(e^{kl} - e^{kx}); \hspace{1cm} p^0_{r-}(x) = Ce^{kl} - D e^{kx}. \label{main:eq:Abs_prob_m_to_0}
\end{equation}
The last relationships can also be obtained by considering absorption at the endpoint $0$ and using the corresponding boundary conditions. It is interesting to note that the escape rate from the paused state, $\mu$, does not influence the probabilities, which is reasonable since it influences neither the direction nor the duration of movement, but only the elapsed time. For later purposes, for the right endpoint, one can write
\begin{equation}
p^l_D(x)= (C-D)e^{kx}; \hspace{1cm} p^l_S(x)= (C+D)e^{kx}-2D,\label{main:eq:pi_sum_diff_l}
\end{equation}
while for the left endpoint
\begin{equation}
p^0_D(x)= (D-C)e^{kx}; \hspace{1cm} p^0_S(x)= 2Ce^{kl} - (C+D)e^{kx}. \label{main:eq:pi_sum_diff_0}
\end{equation}

\subsection{Relationships for MFPTs}

Before deriving the relationships for the MFPTs, it is convenient to introduce the concept of the weighted mean first-passage time (WMFPT). Let $w(x)$ denote the mean time to absorption (MTA) at either endpoint for a walk that starts at position $x$ and in an arbitrary state of motion. As the mean duration of all such paths, it can be defined as 
\begin{equation}
w(x) = \mathbb{E}_x\left[ \int_0^{T_a} dt\right],
\end{equation}
where $T_a$ is the time to absorption at either endpoint,
\begin{equation}
T_a = \min (T_0,T_l).
\end{equation}
For an initial position in the interior of the segment, $T_E=\inf\{t\geq0:X_t=E\}$, $E\in\{0,l\}$, where $X_t$ is the position component of $Y_t$. Quantities for entry at $(0,r+)$ are defined by the one-sided limit $x\downarrow0$ of the interior solutions; the initial entry is not counted as absorption. The analogous convention applies to entry at $(l,r-)$, using $x\uparrow l$.

The WMFPT to a specific endpoint $E$ before reaching the other one, for example, absorption at $l$ before $0$, is
\begin{equation}
w^l(x) = \mathbb{E}_x \left[\int_0^{T_a} \mathbf{1}_{T_l<T_0} dt \right], \label{main:eq:def_MTA}
\end{equation}
where the indicator function $\mathbf{1}_{T_l<T_0}$ is used to separate the paths that are absorbed at $l$ from those that finish at 0. It is weighted since all paths that finish at 0 are counted but have weight zero. If one uses the same idea to define $w^0(x)$, one obtains 
\begin{equation}
w(x) = w^l(x) + w^0(x),
\end{equation}
since $\mathbf{1}_{T_l<T_0} + \mathbf{1}_{T_0<T_l} = 1$. Let us now split the integral at an infinitesimal time $h$ 
\begin{equation}
w^l(x) = \mathbb{E}_x\left[\int_0^h \mathbf{1}_{T_l<T_0} dt + \int_h^{T_a} \mathbf{1}_{T_l<T_0} dt\right].
\end{equation}
The first integral, up to order $o(h)$, gives
\begin{equation}
\mathbb{E}_x\left[\int_0^h \mathbf{1}_{T_l<T_0} dt\right] = p^l(x) h, 
\end{equation}
since the indicator function in the interval $(0, h)$ determines the fraction of the paths that are absorbed at $l$ -- the probability of absorption at $l$, while the rest are absorbed at $0$. This holds because the probability of absorption over a very small time interval $h$ is negligible. For the second integral, one obtains an expectation of the future WMFPTs
\begin{equation}
\mathbb{E}_x\left[\int_h^{T_a} \mathbf{1}_{T_l<T_0} dt\right] = \mathbb{E}_x\left[ w^l(Y_h) \right],
\end{equation}
which leads to
\begin{equation}
w^l(x) = p^l(x) h + \mathbb{E}_x\left[ w^l(Y_h) \right].
\end{equation}
Recall that the future state of the process $Y_h$ can differ in both position and state of motion. Using this relationship in the definition of the generator $\mathcal{L}$ gives
\begin{equation}
\left(\mathcal{L}w^l\right)(x) = \lim_{h\to 0}\frac{\mathbb{E}_x\left[w^l(Y_h) \right] - w^l(x)}{h} = -p^l(x), \label{main:eq:weig_MFPT_gen}
\end{equation}
where the right-hand side represents the forcing term. Note that the same relationship holds for the respective generator for reaching the other endpoint $0$ before $l$. Using the result \eqref{main:eq:weig_MFPT_gen} in the generator equations for an arbitrary test function \eqref{main:eq:L_general} gives the following equations for the WMFPTs
\begin{eqnarray}
(\mathcal{L}w^E_{r+})(x) &=& v\frac{dw^E_{r+}(x)}{dx}+\lambda\left[w^E_{p+}(x)-w^E_{r+}(x)\right] = -p^E_{r+}(x);\nonumber
\\
(\mathcal{L}w^E_{r-})(x) &=& -v\frac{dw^E_{r-}(x)}{dx}+\lambda\left[w^E_{p-}(x)-w^E_{r-}(x)\right] = -p^E_{r-}(x);\nonumber\\
(\mathcal{L}w^E_{p+})(x) &=& \mu\left[\alpha w^E_{r+}(x)+(1-\alpha)w^E_{r-}(x)-w^E_{p+}(x)\right] = -p^E_{p+}(x);\nonumber\\ 
(\mathcal{L}w^E_{p-})(x) &=& \mu\left[(1-\beta) w^E_{r+}(x)+\beta w^E_{r-}(x)-w^E_{p-}(x)\right] = -p^E_{p-}(x). \label{main:eq:L_gen_omega}
\end{eqnarray}
The only difference from the case with the absorption probability as the test function is the appearance of nonzero terms on the right-hand side, so the WMFPT is obtained from a Poisson problem. 

The WMFPT is a useful concept, but the more biologically relevant quantity is the conditional mean first-passage time (CMFPT), $c^E(x)$, which is the {\it mean} time of first arrival at a specified endpoint $E$ before reaching any other endpoint -- thus, only successful trajectories are counted. The two quantities are related by
\begin{equation}
c^E(x) = \frac{w^E(x)}{p^E(x)}, \label{main:eq:conditional_mfpt_segment}
\end{equation}
which follows by comparing the average over successful paths with the weighted average over all paths. 

The calculations of WMFPTs are slightly more involved than those of the absorption probabilities. To proceed, first use the relationships between the absorption probabilities from paused and running states \eqref{main:eq:abs_prob_rest_from_m} in the last two algebraic equations in \eqref{main:eq:L_gen_omega} to obtain
\begin{eqnarray}
w^E_{p+}(x) &=& \alpha w^E_{r+}(x)+(1-\alpha)w^E_{r-}(x) + \frac{\alpha p^E_{r+}(x)+(1-\alpha)p^E_{r-}(x)}{\mu};\nonumber\\ 
w^E_{p-}(x) &=& (1-\beta) w^E_{r+}(x)+\beta w^E_{r-}(x) +\frac{(1-\beta) p^E_{r+}(x)+\beta p^E_{r-}(x)}{\mu}. \label{main:eq:w_rest_from_m}
\end{eqnarray}
Using these relationships and those for the absorption probabilities, the differential equations for the moving states become
\begin{eqnarray}
\frac{dw^E_{r+}(x)}{dx} &=& A w^E_{r+}(x) - A w^E_{r-}(x) - \frac{1}{v}\left[ \rho(\alpha p^E_{r+} + (1-\alpha)p^E_{r-}) + p^E_{r+}\right]; \nonumber \\
\frac{dw^E_{r-}(x)}{dx} &=& B w^E_{r+}(x) - B w^E_{r-}(x) + \frac{1}{v}\left[\rho((1-\beta)p^E_{r+} + \beta p^E_{r-}) + p^E_{r-}\right], \label{main:eq:w_system}
\end{eqnarray}
where $A$ and $B$ are as before and $\rho = \lambda/\mu$. One can proceed with the same approach by considering the difference $w^E_D(x)=w^E_{r+}(x) - w^E_{r-}(x)$ and the sum $w^E_S(x) = w^E_{r+}(x) + w^E_{r-}(x)$. The respective differential equations for the difference and sum are
\begin{eqnarray}
\frac{dw^E_D(x)}{dx} &=& (A-B)w^E_D(x) -\frac{1}{v}\left[\rho(\alpha-\beta)p^E_D(x)+(1+\rho)p^E_S(x) \right]; \nonumber\\
\frac{dw^E_S(x)}{dx} &=& (A+B)w^E_D(x) +\frac{1}{v}\left[\rho(1-\alpha-\beta)- 1\right] p^E_D(x). \label{main:eq:SD_WMFPT_ODE}
\end{eqnarray}
The procedure for solving the system of ODEs is rather technical and similar to that for the absorption probabilities. The remaining details for obtaining the specific forms of the solutions are provided in the Supplemental Material \cite{SM}. 

It is interesting to note that although the WMFPTs have different asymptotic behavior depending on the relative values of $\alpha$ and $\beta \neq \alpha$, the CMFPTs grow linearly with the length of the dendrite $l$ in both cases. The CMFPTs differ in their asymptotic slopes, which depend on the relative values of the persistence probabilities $\alpha$ and $\beta$. The derivation of the asymptotic behavior for both scenarios, $\alpha > \beta$ and $\beta > \alpha$, is provided in the Supplemental Material \cite{SM}. The linear asymptotes for the CMFPTs are given by
\begin{eqnarray}
c^l_{\beta>\alpha} &\approx & \frac{(1+\rho)(2-\alpha-\beta)-\rho(\beta-\alpha)^2}{v(\beta-\alpha)}l=S_{\rm ret} l; \nonumber \\
c^l_{\alpha>\beta} &\approx & \frac{(1+\rho)(2-\alpha-\beta)}{v(\alpha-\beta)}l=S_{\rm ant} l, \label{eq:CMFPT_lin_slope}
\end{eqnarray}
where $S_{\rm ant}$ and $S_{\rm ret}$ are the respective slopes of the asymptotic growth. The negative term in the first of the last two relationships indicates that, when $\beta > \alpha$, the slope of the asymptote can be smaller than in the opposite case, $\alpha > \beta$. Obviously, when $\alpha$ is much larger than $\beta$ -- corresponding to strong anterograde persistence -- mRNA transport greatly benefits from frequent anterograde runs and arrival at the opposite end is faster.

\subsection{Semi-Markov model of motion throughout the dendritic tree}

Before using the segment-level results for the whole neuron, let us first consider the general case of a semi-Markov chain on a dendritic tree. A semi-Markov chain is a generalization of an ordinary Markov chain in which the sojourn times at each state are not necessarily exponentially distributed. Consider a chain with $N$ states, $i=1,2,\dots,N$, transition probabilities $p_{ij}$ between states $i$ and $j$, and conditional waiting-time densities $C_{ij}(\tau)$. The probabilities $p_{ij}$ are constants and define the embedded Markov chain with transition matrix $\mathbf{P}$. The waiting-time distributions are conditioned on both $i$ and $j$ and may differ between state pairs; their cumulative distribution functions are denoted by $F_{ij}(t)$. The corresponding mean sojourn times are
\begin{equation}
\mathbb{E}[\mathcal{C}_{ij}] = \int_0^{\infty} \tau C_{ij}(\tau) d\tau = c_{ij}. \label{eq:Cond_mean_wait_def}
\end{equation}
The Markov property assumes that transition probabilities and waiting times are independent of past states, which is useful for calculating multistep transitions. Consider a random walk along an {\it absorbed path} that starts at state $i$ and stops upon first arrival at an absorbing state (synapse $s$). Such a path is a random sequence of states $\hat{\sigma}_{is}=\{k_0=i, k_1, k_2, \dots k_{n-1}, k_n=s;k_m \neq s, m<n\}$. These paths constitute the sets $\Sigma_{is}$ for each pair consisting of a starting state $i$ and a target $s$. The probability of a specific path $\sigma_{is}$ is the product of its one-step transition probabilities
\begin{equation}
P(\sigma_{is}) = p_{k_0k_1}p_{k_1k_2}\dots p_{k_{n-1}k_n}. \label{eq:Path_prob_decomp}
\end{equation}
The memoryless property implies that the mean value of the total travel time along that specific path is
\begin{equation}
c_{\sigma_{is}} = \mathbb{E}\left[\mathcal{C}_{is}|\sigma_{is}\right] = \mathbb{E}\left[\sum_{m=0}^{n-1} \mathcal{C}_{k_mk_{m+1}}\right] = \sum_{m=0}^{n-1} c_{k_mk_{m+1}}, \label{eq:MWT_along_path}
\end{equation}
where $\mathcal{C}_{ij}$ is the random waiting time from node $i$ to $j$ with mean $\mathbb{E}[\mathcal{C}_{ij}] = c_{ij}$. Now the CMFPT from a given starting node $i$ to a specified target $s$ (conditioned on absorption at $s$ and not elsewhere) can be calculated as the corresponding mean over all possible paths starting from $i$ and absorbed at $s$
\begin{equation}
\gamma_{is} = \mathbb{E}\left[c_{\hat{\sigma}_{is}}\right] = \mathbb{E}\left[\mathbb{E}\left[\mathcal{C}_{is}|\hat{\sigma}_{is}\right]\right],
\end{equation}
As will be seen below, it is more convenient to calculate the associated WMFPT, where the mean time for each specific path $\sigma_{is}$ is weighted by the probability of that path occurring in the random walk 
\begin{equation}
\omega_{is} = \sum_{\sigma_{is} \in \Sigma_{is}} P(\sigma_{is})c_{\sigma_{is}} = \gamma_{is}\pi_{is}, \label{eq:MFPT_prod_sum}
\end{equation}
where
\begin{equation}
\pi_{is} = \sum_{\sigma_{is} \in \Sigma_{is}} P(\sigma_{is})
\end{equation}
is the probability of absorption at $s$, starting from $i$. As for the case of transport along a dendrite, the last relationship suggests that the CMFPT and WMFPT are related by
\begin{equation}
\gamma_{is} = \frac{\omega_{is}}{\pi_{is}}, \label{eq:MFPTs_relationship}
\end{equation}
Before deriving a relationship for calculating the WMFPT \eqref{eq:MFPT_prod_sum}, let us assume that there might be more than one synapse $s$. For problems with multiple absorbing states, one conveniently applies the classical theory of absorbing Markov chains. Then the transition probability matrix $\mathbf{P}$ is organized in the canonical form \cite{grinstead2012introduction}
\begin{equation}
\mathbf{P} = \begin{bmatrix}
\mathbf{Q} & \mathbf{S}\\
\mathbf{0} & \mathbf{I}
\end{bmatrix}, \label{eq:CAN_Form}
\end{equation}
where $\mathbf{Q}$ is the submatrix that corresponds to transitions between transient states, $\mathbf{S}$ accounts for the transitions to absorbing states, and $\mathbf{I}$ is an identity matrix of appropriate size. To conveniently combine the transition probabilities $p_{ij}$ with the respective one-step CMFPTs $c_{ij}$ \eqref{eq:MWT_along_path}, it is useful to construct an auxiliary matrix $\mathbf{A}$ with elements 
\begin{equation}
a_{ij} = p_{ij}e^{rc_{ij}},\label{eq:Auxil_A_elem}
\end{equation}
where $r$ is an auxiliary variable. The elements of the product $\mathbf{A}^2$ of the matrix with itself, $a^{(2)}_{ij}$, are sums of the probabilities of all two-step paths from $i$ to $j$ multiplied by the exponentials that keep track of the mean time spent along each such path
\begin{equation}
a^{(2)}_{ij} = \sum_{k=1}^N p_{ik}p_{kj} e^{r(c_{ik}+c_{kj})}.
\end{equation}
Thus, to account for both the probability that a walker starting from $i$ will be at $j$ after two steps and the mean time required, one can use 
\begin{equation}
\frac{\partial a^{(2)}_{ij}}{\partial r}\Big|_{r=0} = \sum_{k=1}^N p_{ik}p_{kj}(c_{ik}+c_{kj}) = \sum_{\substack{\sigma_{ij}\in \Sigma_{ij}\\ L(\sigma_{ij}) = 2}} P(\sigma_{ij})c_{\sigma_{ij}}.
\end{equation}
where the function $L$ determines the length of the path $\sigma_{ij}$ in terms of the number of steps. Note that $\sigma_{ij}$ is not a path that necessarily makes its first visit to $j$ in the last step, and $\Sigma_{ij}$ is the set of all paths that start at $i$ and finish at $j$. By induction, one can conclude that the element of the $n$-th power $a^{(n)}_{ij}$ contains information about the probability that a walker starting from $i$ will, after $n$ steps, be at $j$ and the average time spent on such journeys. Thus one has
\begin{equation}
\frac{\partial a_{ij}^{(n)}}{\partial r}\Big|_{r=0} = \sum_{\substack{\sigma_{ij}\in \Sigma_{ij}\\ L(\sigma_{ij}) = n}} P(\sigma_{ij})c_{\sigma_{ij}}. \label{eq:A_n_th_deriv}
\end{equation}
Absorbing states $s$ are such that the probability of exit is zero, $p_{sj}=0; j\neq s$, so each path that ends at $s$ makes its first visit to $s$ in the last step. Now, one can organize the possible first-passage paths from $i$ to $s$ by their length and write the WMFPT \eqref{eq:MFPT_prod_sum} as 
\begin{equation}
\omega_{is} = \sum_{n=1}^{\infty} \sum_{\substack{\sigma_{is}\in \Sigma_{is}\\ L(\sigma_{is}) = n}} P(\sigma_{is})c_{\sigma_{is}} = \sum_{n=1}^{\infty}\frac{\partial a^{(n)}_{is}}{\partial r}\Big|_{r=0}. \label{eq:MFPT_sum_by_length}
\end{equation} 
By the linearity of differentiation,
\begin{equation}
\omega_{is} =\frac{\partial}{\partial r}\left[ \sum_{n=1}^{\infty} a_{is}^{(n)}\right]\Big|_{r=0} = \frac{\partial}{\partial r}\left[ \sum_{n=1}^{\infty}\sum_{k \in \mathcal{T}} a_{ik}^{(n-1)}a_{ks}\right]\Big|_{r=0}
\end{equation}
where the last step, which corresponds to absorption, is separated. This step is needed in order to have a substochastic matrix, as will be seen below. For more compact notation, one can collect the WMFPTs from all other nodes $i$ to the target $s$ in a vector $\bm{\omega}_s$. Then one has the relationship
\begin{equation}
\bm{\omega}_s = \frac{\partial}{\partial r}\left[\sum_{n=0}^{\infty}\mathbf{A}_{\rm tt}^{n}\mathbf{a}_s\right]\Big|_{r=0} = \frac{\partial}{\partial r}\left[\left(\mathbf{I}-\mathbf{A}_{\rm tt}\right)^{-1}\mathbf{a}_s\right]\Big|_{r=0} \label{eq:MFPT_j}
\end{equation}
where $\mathbf{A}_{\rm tt}$ is the submatrix of $\mathbf{A}$ that corresponds to the transitions between transient states, while the vector $\mathbf{a}_s$ has elements $a_{js}$. At \(r=0\), \(\mathbf{A}_{tt}(0)=\mathbf{Q}\) is substochastic and, for an absorbing chain, \(\rho(\mathbf{Q})<1\). Hence the matrix series converges in a neighborhood of \(r=0\), where the derivative is evaluated, yielding \((\mathbf{I}-\mathbf{A}_{tt})^{-1}\). Let $\mathbf{A}_{\rm ta}$ denote the submatrix of $\mathbf{A}$ associated with the transitions from transient to absorbing states. Since one can write a corresponding vector \eqref{eq:MFPT_j} of WMFPTs for each absorbing state $s$, all such vectors can be used as columns of a matrix $\bm{\Omega}$ of WMFPTs from all transient to all absorbing states. The WMFPT matrix will be 
\begin{equation}
\bm{\Omega} = \frac{\partial}{\partial r}\left[\left(\mathbf{I}-\mathbf{A}_{\rm tt}\right)^{-1}\mathbf{A}_{\rm ta}\right]\Big|_{r=0}. \label{main:eq:MFPT_all}
\end{equation}
By evaluating the derivative in \eqref{main:eq:MFPT_all} and making extensive use of the matrix algebra detailed in the Supplemental Material \cite{SM}, one obtains the final expression for the WMFPT matrix $\mathbf{\Omega}$, which has a compact form
\begin{equation}
\mathbf{\Omega} = \mathbf{ND},
\end{equation}
where $\mathbf{N}$ is the fundamental matrix of the embedded Markov chain, while $\mathbf{D}$ is the matrix of appropriately weighted sojourn times along dendrites. The matrix element $n_{ij}$ represents the expected number of visits to node $j$ before absorption (at an arbitrary absorbing state) by a walker that started at $i$. The element of the weighted sojourn time matrix
\begin{equation}
d_{ks} = p_{ks} c_{ks} + \sum_{m \in \mathcal{T}} p_{km} c_{km} \pi_{ms}, \label{eq:MWT_matr_elem}
\end{equation}
represents the one-step time contribution from transient node $k$ for a walk that is {\it absorbed} at a specified target $s$. The details of this derivation are given in the Supplemental Material \cite{SM}. Fig.~\ref{fig:MFPT_illustrated} provides a visual interpretation of the meaning of the elements $d_{ks}$. The widths of the colored tubes in the figure are proportional to the probability of absorption at a given synapse.

To complete the model, the results for transport along a single dendritic segment, namely the absorption probabilities and CMFPTs, are incorporated into the semi-Markov model. The CMFPTs along dendrites are used as the mean sojourn times $c_{ij}$ in \eqref{eq:Cond_mean_wait_def}, while the absorption probabilities are used to construct the transition matrix $\mathbf{P}$. 

Transport throughout the whole neuron, from the soma to an arbitrary synapse, can be analyzed by constructing a coarse-grained tree network model. The branching points in the neuron, along with the soma and synapses, become nodes in the model network, while the dendritic segments become edges. The attempt to transition from node $i$ to $j$ is characterized by the probability of choosing the dendritic segment towards $j$, which, for simplicity, is assumed to be uniform, $1/d_i$, where $d_i$ is the number of dendrites that meet at $i$. Note that this is likely an oversimplification, since it allows a return to the dendritic segment from which the mRNA arrived and the choice probabilities do not depend on dendrite widths and the number of microtubules they contain. The probability of successfully arriving at node $j$ instead of returning to the start $i$ is thus $p_{ij} = p^l_{ij}/d_i$, where $p^l_{ij}$ is the dendrite traversal probability. The latter can be either $p^{l}_{ij}(0)$ or $\overline{p}^{l}_{ij}(0)$, depending on the direction from $i$ to $j$ (anterograde or retrograde), which requires the appropriate use of $\alpha$ and $\beta$ for the persistence. The probability of returning to the same node without reaching the other end accounts for all unsuccessful attempts and is obtained from normalization
\begin{equation}
p_{ii} = 1 - \sum_{k=1}^{d_i}p_{ik} = \frac{1}{d_i}\sum_{k=1}^{d_i}(1-p^l_{ik}).
\end{equation}
The sojourn time in the semi-Markov chain for the transition $i \to j$ is simply the CMFPT for passing along the dendrite in the respective direction. Accounting for all returns and the respective CMFPTs gives the sojourn time for the loop $i \to i$
\begin{equation}
c_{ii} = \frac{1}{d_i}\sum_{k=1}^{d_i}c_{ii}^{(k)},
\end{equation}
where the superscript $(k)$ is used to denote the time associated with an unsuccessful attempt to reach node $k$, that is, $c^{0}_{ik}(0)$ or $\overline{c}^{0}_{ik}(0)$ depending on whether the direction $i \to k$ is anterograde or retrograde. As a visual aid, Fig.~\ref{fig:model}(c) provides a small toy model of a neuron with all possible transitions and major parameters shown for one branching node.

\begin{figure}[t]
\centering
\includegraphics[width=\columnwidth]{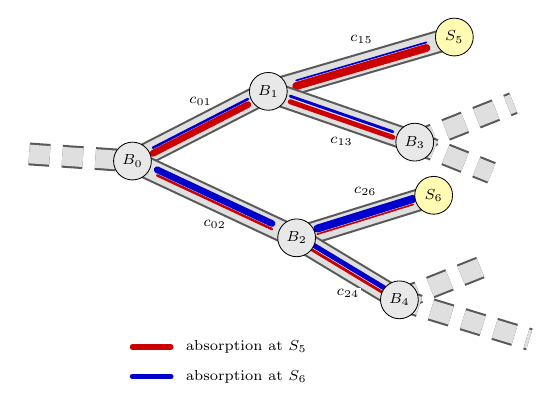}
\caption{Illustration of the contributions of the transition probabilities and waiting times along dendritic segments to the calculation of the CMFPT to synapses, as given in \eqref{eq:MWT_matr_elem}. The widths of the colored stripes are proportional to the absorption probabilities, or the fractions of trajectories absorbed at synapse $S_5$ (red) or $S_6$ (blue) that pass through a given dendritic segment. Retrograde transitions and related quantities are not shown for simplicity.}
\label{fig:MFPT_illustrated}
\end{figure}

\subsection{Asymptotic regimes}

Here we provide the asymptotic relationships for the CMFPTs throughout the whole neuron, which hold for very long dendrites, $|k|l\gg 1$. The details of the derivation are provided in the Supplemental Material \cite{SM}.

\subsubsection{Retrograde dominance regime}

In the present terminal-target representation, each synaptic target $s$ is connected to the transient dendritic network through a single neighboring branching point $b$, so only one $p_{ks}$ or $c_{ks}$ is nonzero. Further, for rather long dendrites when $\beta>\alpha$, the probability of mRNA reaching the opposite end in the anterograde direction is very small. Then the cargo stays in the transient subnetwork for a very long period. This is verified by the observation that the leading eigenvalue $\Lambda_1$ of the transient submatrix $\mathbf{Q}$ is very close to unity. The second eigenvalue, $\Lambda_2$, is also close to one, but considerably farther from it, $1-\Lambda_1 \ll 1 - \Lambda_2$, so regardless of the starting node (soma or branching point), the random walk of the mRNA can be considered to be in a quasi-stationary state for a relatively long period before absorption. Thus, one might consider that the rows of the fundamental matrix $\mathbf{N}$ are nearly identical, and write $n_{km} \approx \tilde{n}_m$, where the tilde is used to denote that the expected number of visits does not depend on the starting node. This implies that the probability of absorption at a given synapse might be written as
\begin{equation}
\pi_{ks} = \sum_{m} n_{km}p_{ms} \approx \tilde{n}_b p_{bs},
\end{equation}
since the synapse $s$ is accessible only through its nearest branching node $b$. Then, the waiting-time matrix elements are
\begin{equation}
d_{ks} = p_{ks} c_{ks} + \sum_{m} p_{km} c_{km} \pi_{ms} = \delta_{kb} p_{ks} c_{ks} + \tilde{n}_b p_{bs} \sum_{m} p_{km} c_{km},
\end{equation}
where the Kronecker symbol $\delta_{kb}$ indicates that the term $p_{ks} c_{ks}$ is nonzero only for the nearest branching point $k = b$. Thus, the WMFPT from the soma to the synapse $s$ is
\begin{eqnarray}
\omega_{Ss} = \sum_{k} \tilde{n}_k d_{ks} &=& \sum_{k \neq b} \tilde{n}_k \left(\tilde{n}_b p_{bs}  \sum_{m} p_{km} c_{km} \right) + \tilde{n}_b \left( p_{bs} c_{bs} + \tilde{n}_b p_{bs} \sum_{m} p_{bm} c_{bm} \right) \nonumber \\
&=& \tilde{n}_b p_{bs} \left( \sum_{k} \tilde{n}_k \sum_{m\neq s} p_{km} c_{km} + c_{bs}  \right),
\end{eqnarray}
where the capital $S$ is the index of the soma. Then the CMFPT to the synapse $s$ is
\begin{equation}
\gamma_{Ss} = \frac{\omega_{Ss}}{\pi_{Ss}} =   \sum_{k} \tilde{n}_k \sum_{m\neq s} p_{km} c_{km} + c_{bs} \approx \sum_{k} \tilde{n}_k \sum_{m\neq s} p_{km} c_{km},
\end{equation}
since the double sum is much larger than the waiting time $c_{bs}$. The last result has a very simple interpretation. It accounts for the expected number of visits to each branching node before absorption $\tilde{n}_k$ and the expected time spent reaching its neighbors in each step $\sum_{m\neq s} p_{km} c_{km}$. This total time dominates the CMFPT, while the neglected term $c_{bs}$ corresponds to the last step that leads to absorption. Thus, in this regime, the CMFPT is nearly equal for all synapses and is determined by the time spent in the transient subnetwork of branching points. The CMFPT is thus independent of the distance from the soma to the synapse and is determined by the structure of the dendritic tree and the transition probabilities between branching points. 

\subsubsection{Anterograde dominance regime}

When anterograde motion is favored, $\alpha > \beta$, for long dendritic segments, $|k|l \gg 1$, the probability of reaching the other end in the anterograde direction is $p_{r+}^l(0)\approx 1 - B/A = (\alpha - \beta)/(1 - \beta)$. In this case, the probability of reaching the ``repelling'' end of a dendritic segment (retrograde motion) vanishes, and the mRNA almost surely returns to the branching point if it left in the retrograde direction. Since three dendrites meet at a branching point, the transition probability from a branching point toward other branching points or synapses in the anterograde direction is $(1/3)\cdot(\alpha - \beta)/(1 - \beta)$. The dendritic segment leading in the retrograde direction has an exponentially vanishing traversal probability, so the probability of a loop -- an unsuccessful trip of the mRNA cargo along any dendritic segment -- is $1 - (2/3)\cdot(\alpha - \beta)/(1 - \beta)$. 

The vanishing probability of successful retrograde traversal makes the coarse-grained motion nearly unidirectional -- anterograde. Then the CMFPT to a synapse is the sum of the CMFPTs for each dendrite along the unique path from the soma to the synapse. The CMFPT to a synapse $s$ therefore has the following general form
\begin{equation}
    \gamma_{Ss} = \sum_{\mathrm{edge}\in \mathcal{P}_{Ss}} c_{\mathrm{edge}}(l_{\mathrm{edge}}) + \mathrm{O}(n_{Ss}),
\end{equation}
where $\mathcal{P}_{Ss}$ is the set of edges along the unique path from the soma to the synapse $s$, and $c_{\mathrm{edge}}(l_{\mathrm{edge}})$ is the CMFPT along each edge, which depends on the length of the dendrite $l_{\mathrm{edge}}$. The $\mathrm{O}(n_{Ss})$ term accounts for the finite time spent at branching points, which becomes negligible for long dendrites. Since, for long dendrites, the CMFPT along each edge grows linearly with the length of the dendrite as $c_{\mathrm{edge}}\approx S_{\mathrm{ant}}l_{\mathrm{edge}}$, one has a simple relationship for the CMFPT to a synapse in this regime
\begin{equation}
    \gamma_{Ss} \approx S_{\mathrm{ant}} \sum_{\mathrm{edge}\in \mathcal{P}_{Ss}} l_{\mathrm{edge}} = S_{\mathrm{ant}} l_{Ss}.
\end{equation}
The detailed derivation of the last result is provided in the Supplemental Material \cite{SM}. Thus, in the anterograde-dominant regime, the leading contribution to the CMFPT is proportional to the total dendritic length along the unique soma–synapse path.

\section{Numerical examples}

\begin{figure}[t]
\includegraphics[width=\columnwidth]{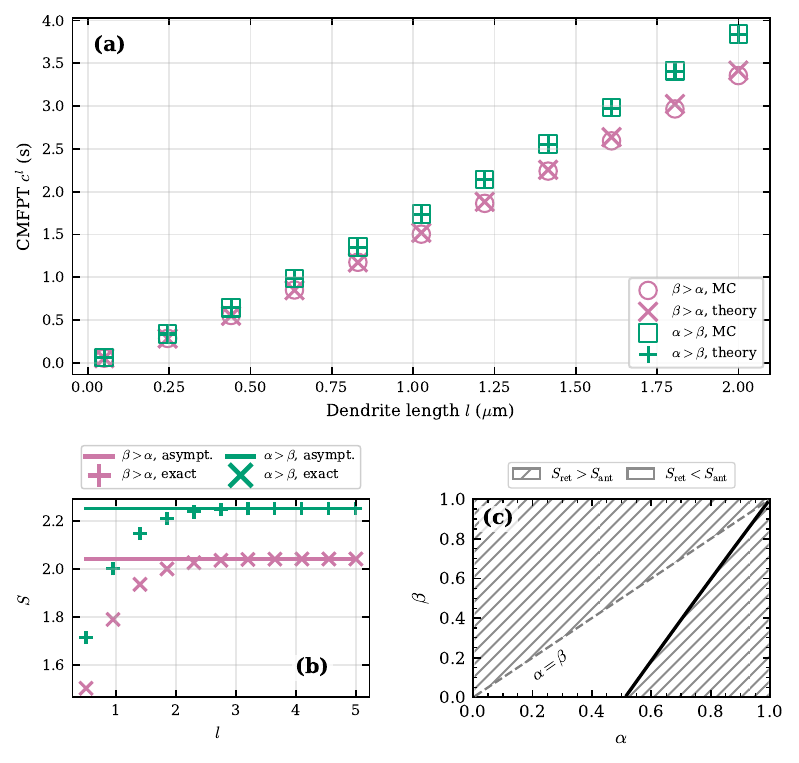}
\caption{(a) Comparison of the exact theoretical CMFPT along a dendrite with Monte Carlo simulations for $(\alpha, \beta) \in \{(0.2, 0.8), (0.8, 0.2)\}$. (b) Convergence towards the asymptotic value (for $l\to\infty$) of the slopes of the CMFPT curves as a function of $l$. (c) Regions in the $\alpha-\beta$ parameter space where one slope of the linear growth is greater than the other. On the dashed boundary line $\alpha=\beta$, the growth is quadratic.}
\label{fig:telegraph_asymp_beh}
\end{figure}

To gain insight into the dependence of the CMFPT on the variables that determine it, the parameter values were taken from the detailed study of mRNA transport \cite{song2018neuronal}. The rate parameters are $\lambda = 4.4 s^{-1}$ and $\mu = 12.5 s^{-1}$, while the velocity is $v=1.0 \mu m/s$, which is close to the values obtained in other studies \cite{fusco2003single, kiebler2006neuronal, amrute2012single}. The first analysis presented here examines the influence of the persistence probabilities $\alpha$ and $\beta$ on the conditional mean time to traverse a dendritic segment. Fig.~\ref{fig:telegraph_asymp_beh} shows Monte Carlo simulations of mRNA motion along a relatively short dendritic segment of length $l = 5.0 \mu m$, along with the theoretical predictions. The persistence probability values (0.2 and 0.8) are chosen to differ substantially from each other and from those estimated from the observations in order to obtain better separation of the curves. 
A counterintuitive observation is that when retrograde persistence is stronger than anterograde persistence ($\beta > \alpha$), the conditional mean travel time away from the soma is slightly shorter than the traversal time in the opposite direction. The corresponding probability of successful arrival is, however, exponentially smaller. The exact expression for the CMFPT is not convenient for analyzing this phenomenon, but the asymptotic expression can be used for very long dendritic segments, $|k|l\gg 1$ (see Fig.~\ref{fig:telegraph_asymp_beh}(b) for asymptotics). As shown in the previous section, the CMFPT grows linearly with the length of the dendrite, and the slope of the linear growth is given by \eqref{eq:CMFPT_lin_slope}. The slope for the case when retrograde persistence is more pronounced is smaller than that for the opposite case. This means that for long enough dendrites, the conditional mean time to reach the other end is shorter when retrograde persistence is stronger. The regions where the CMFPT is smaller when retrograde persistence is stronger are shown in Fig.~\ref{fig:telegraph_asymp_beh}(c).

\begin{figure*}
\begin{minipage}{0.45\textwidth}
\begin{overpic}[width=\linewidth]{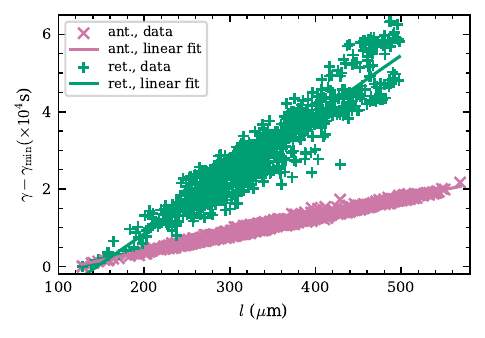}
    \put(5,70){\textbf{(a)}}
\end{overpic}
\end{minipage}
\hfill
\begin{minipage}{0.45\textwidth}
\begin{overpic}[width=\linewidth]{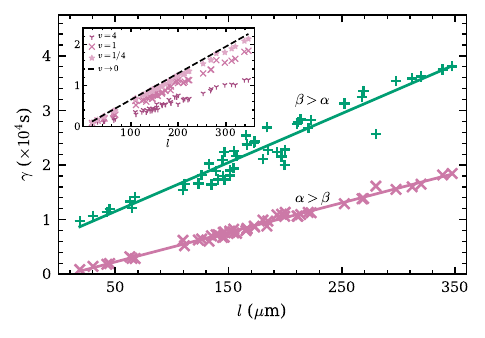}
    \put(5,70){\textbf{(b)}}
\end{overpic}
\end{minipage}
\vspace{0.4cm}
\caption{(a) CMFPT for mRNA cargo transport from the soma to a synapse as a function of the distance between them for a human neuron  \cite{masoli2024human}. Note that the plot shows the difference between the actual and minimum values of the CMFPT. (b) The same plot for a pyramidal hippocampal neuron of a mouse \cite{benavides2020differential}. The magenta crosses represent calculations for stronger anterograde persistence ($\alpha=0.53$ and $\beta=0.51$), while the green plus signs represent stronger retrograde persistence ($\alpha = 0.51; \beta = 0.53$). The inset in (b) shows the dependence of the CMFPT for different velocities in the running phase, which effectively correspond to different dendritic segment lengths. Convergence to the theoretical line indicates that random-walk transport in a neuron resembles a persistent random walk on a line.}
\label{fig:MFPT_neuron}
\end{figure*}

Next, the results for transition probabilities and CMFPTs for transport along dendritic segments are applied in a semi-Markov model of a neuron. Random mRNA transport from the soma (or the root node in neurons for which the soma is not identified) to synapses was examined for several neurons, two of which are presented here: a human Purkinje neuron, which is known for its size and dendritic complexity \cite{masoli2024human}, and a mouse pyramidal neuron \cite{benavides2020differential}. The data for all neuron morphologies were obtained from the freely available database NeuroMorpho.Org \cite{Ascoli2007NeuroMorpho}. The neuronal reconstructions used here provide dendritic morphology but do not specify the locations of individual synapses. Therefore, for the numerical implementation, terminal dendritic tips are used as representative synaptic targets. This choice is a modeling approximation and should not be interpreted as implying that synapses are restricted to dendritic terminals. The persistence parameter values were chosen as estimated in \citet{song2018neuronal}, $\alpha=0.51;\beta=0.53$, which favor retrograde motion, together with the corresponding reversed combination $\alpha=0.53;\beta=0.51$. It is interesting that the CMFPT seems to be linearly dependent on the distance between the soma and a synapse. The constant term in the linear fit appears to vanish for stronger anterograde persistence, while it is significant for the opposite scenario, even dwarfing the linearly changing term in some cases. Fig.~\ref{fig:MFPT_neuron} shows the CMFPTs for the human and mouse neurons, revealing two notable features. When retrograde motion is favored, the bias observed for the human neuron is about $6.3\cdot10^7 \rm{s}$, which is three orders of magnitude larger than the differences between CMFPTs for different synapses. This means that for long enough dendrites, $l \gg v / \lambda$, as is the case with this neuron, the CMFPT becomes nearly equal for all synapses. A possible biological interpretation is that successfully delivered mRNA molecules reach different synaptic targets with comparable mean ages. The implications of this result for transcript integrity or translational competence would require an explicit model of mRNA degradation. This observation can be attributed to the long transient period in the semi-Markov chain, during which the walk remains in a quasi-stationary state. This is indeed verified by observing that for the leading eigenvalues of the transient submatrix, it holds $1-\Lambda_1 \ll 1-\Lambda_2$, which implies a long residence in a quasi-stationary state \cite{DarrochSeneta1965}.

An important implication of the retrograde-dominated regime concerns the accessibility of distal synaptic targets. Although the conditional mean first-passage time of successfully delivered cargo becomes only weakly dependent on soma-to-synapse distance, the corresponding absorption probability decreases strongly, approximately exponentially, with distance. Thus, in this regime, the principal limitation of long-range delivery is not the time required to reach a distant target once a successful trajectory is realized, but the small probability of realizing such a trajectory in the first place. A related loss of accessibility of distant targets has been reported in intermittent-search models on dendritic trees \cite{newby2009directed}, although the mechanism considered here arises from directional-persistence asymmetry.

This observation is particularly relevant because the experimentally motivated transport parameters considered here place the system in, or close to, the retrograde-dominated regime. If such a directional bias persists over long dendritic distances, the resulting low probability of reaching distal targets raises the question of how efficient mRNA localization at remote synaptic regions is achieved in vivo. Possible mechanisms include activity-dependent changes in cargo mobility or directional bias, remobilization of previously stationary RNA, local regulation of motor activity, and repeated transport attempts by multiple mRNA molecules. More generally, the present results suggest that the relative strength of anterograde and retrograde persistence may be an important regulatory parameter of dendritic mRNA localization rather than a fixed property of the transport process.

\section{Conclusions}

We developed a two-level analytical framework for studying mRNA transport along microtubules throughout neuronal dendritic trees. At the microscopic level, the PTPP provides a natural description of the experimentally observed run-and-pause dynamics of mRNA cargo along individual dendritic segments, including directional persistence. By coarse-graining these segment-level dynamics, transport at the whole-neuron level is represented as a semi-Markov random walk on the dendritic tree, with dendritic segments constituting the network links. For the PTPP on a single dendritic segment, the symmetric case \(\alpha=\beta\) exhibits diffusion-like quadratic growth of the CMFPT with distance, whereas persistence asymmetry leads to linear long-distance scaling, consistent with the diffusive long-time behavior of the symmetric telegraph process \cite{goldstein1951diffusion,masoliver2017continuous}. A key finding at the single-segment level is that stronger persistence in a given direction does not necessarily imply a shorter conditional traversal time in that direction. For certain parameter values, including the persistence values estimated from experimental data in \citet{song2018neuronal}, the PTPP predicts a shorter conditional transport time in the direction with weaker persistence.

At the whole-neuron level, a biologically important finding is the emergence of two distinct long-distance transport regimes. On the one hand, an anterograde persistence bias makes the coarse-grained transport increasingly directed, with the leading contribution to the CMFPT growing approximately linearly with the soma-to-target path distance. On the other hand, under a retrograde persistence bias, prolonged residence in the transient dendritic network can render the CMFPT nearly independent of the target distance. 

The present coarse-grained implementation treats terminal dendritic tips as representative synaptic targets, modeled as absorbing nodes of the dendritic tree. This choice is made for the numerical implementation and is not intended to imply that synapses occur exclusively at dendritic terminals. The assumption can be relaxed without altering the basic framework: synaptic targets located within dendritic branches can be introduced as additional absorbing nodes at their corresponding arc-length positions, thereby subdividing the affected dendritic segments. 

The framework can be further refined as additional experimental information becomes available. Such refinements may include a more detailed representation of neuronal geometry, nonuniform branching probabilities, or the influence of molecular regulators and local cellular factors on cargo mobility \cite{espadas2024synaptically}. The present model also neglects mRNA degradation, which can be characterized through transcript half-lives \cite{schwanhausser2011global}. An important extension would therefore be to compare transport times to synaptic regions with the lifetime of the transported transcripts, or to incorporate degradation directly into the stochastic transport model. Beyond their role in determining the CMFPTs, the absorption probabilities could also be used to investigate the spatial distribution of localized mRNA as a function of distance from the soma, providing a complementary theoretical perspective to recent experimental analyses of dendritic mRNA distributions \cite{kim2024spatial}.

A further challenge is to distinguish between arrival at a synaptic region and actual localization there. The present model treats target arrival as absorption, whereas in reality an mRNA molecule reaching a synaptic region may be captured with a probability smaller than one, depending on local cellular conditions. Introducing a target-specific capture probability would therefore provide a natural extension of the present framework and could enable a more realistic prediction of the spatial distribution of localized mRNA along dendrites. Such an extension could provide a direct theoretical link between stochastic transport dynamics and experimentally measured spatial profiles of dendritic mRNA \cite{kim2024spatial}.

Although motivated by intracellular mRNA transport, the proposed framework may also apply to a broader class of stochastic transport problems on filamentous and network-like structures. Potential examples include intracellular transport of other molecular cargoes and organelles \cite{bressloff2013stochastic}, protein sliding along DNA \cite{berg1981diffusion}, and molecular transport through channels and pores \cite{bauer2006molecular}.

\FloatBarrier
\section*{Statements and Declarations}
\subsection*{Acknowledgement}
The author thanks Erin Schuman and Jenna Wingfield for valuable discussions and insightful comments concerning dendritic RNA transport and the biological relevance of first-arrival processes. ChatGPT (OpenAI) was used during the development of this study as an interactive research and learning aid to explore potential applications of the author’s pre-existing semi-Markov framework and alternative stochastic-process formulations. It assisted in identifying the Persistent Telegraph Process with Pauses as a suitable microscopic modeling framework and in exploring a generator-based conditioned-process approach for deriving first-passage quantities. It was also used to assist with intermediate analytical derivations, computational code implementing the mathematical expressions, and subsequent language editing and technical manuscript preparation. The author independently worked through, checked, and validated the final mathematical derivations, computational implementation, numerical results, scientific interpretations, and conclusions, and takes full responsibility for the content of the manuscript. 
\subsection*{Funding}
The author acknowledges the financial support provided by the Faculty of Computer Science and Engineering, Ss. Cyril and Methodius University in Skopje, Macedonia.
\subsection*{Competing Interests}
The author has no relevant financial or non-financial interests to disclose.
\subsection*{Data Availability}
The neuronal morphology data used in this study are publicly available from NeuroMorpho.Org. The code used for the simulations and theoretical analysis is available at the GitHub repository in Ref.~\cite{BasnarkovCode}.

\section*{Supplementary Information}
Supplementary theoretical derivations, notation tables, and additional numerical results are included at the end of this document.

\bibliographystyle{sn-basic}
\bibliography{NeuronBIB}
\clearpage
\appendix
\setcounter{equation}{0}
\setcounter{figure}{0}
\setcounter{table}{0}
\renewcommand{\theequation}{S\arabic{equation}}
\renewcommand{\thefigure}{S\arabic{figure}}
\renewcommand{\thetable}{S\arabic{table}}
\renewcommand{\theHequation}{S\arabic{equation}}
\renewcommand{\theHfigure}{S\arabic{figure}}
\renewcommand{\theHtable}{S\arabic{table}}
\begin{center}
{\LARGE\bfseries Supplementary Information\par}
\vspace{0.4cm}
{\large\bfseries A Persistent Random-Walk Model of Molecular Transport in Neuronal Dendritic Trees\par}
\vspace{0.3cm}
Lasko Basnarkov
\end{center}
\input{supplement_body.tex}
\FloatBarrier
\end{document}

%% file: preamble.tex
\documentclass[11pt,a4paper]{article}
\usepackage[margin=1in]{geometry}
\usepackage[T1]{fontenc}
\usepackage[utf8]{inputenc}
\usepackage{amsmath,amssymb,graphicx}
\usepackage{longtable,booktabs,tabularx}
\usepackage[percent]{overpic}
\usepackage{dcolumn,bm,xcolor}
\usepackage{placeins}
\usepackage[sort,authoryear]{natbib}
\usepackage[hyperfootnotes=false,hidelinks]{hyperref}
\setcitestyle{authoryear,round,semicolon,aysep={},yysep={,}}
\let\cite\citep

\newcommand{\notationrow}[2]{\shortstack[l]{#1} & #2 \\\\}
\graphicspath{{figures/}}

%% file: supplement_body.tex
This document accompanies the main text and provides details of the theoretical analysis, including definitions, derivations of the main results, and analyses of special cases. It also includes a table of the notation used in both the main text and this supplement. The theoretical material is organized into two main sections: one devoted to the Persistent Telegraph Process with Pauses (PTPP) on a single dendritic segment, and the other to the semi-Markov model of stochastic motion throughout the neuron. The final section describes the numerical simulations and presents results showing the dependence of the conditional mean first-passage time (CMFPT) on soma-to-synapse distance for several additional reconstructed neuronal morphologies.

\begin{table}[ht]
\centering
\caption{List of abbreviations.}
\label{tab:abbreviations}
\begin{tabular}{ll}
\hline
\textbf{Abbreviation} & \textbf{Meaning} \\
\hline
CMFPT & Conditional mean first-passage time \\
CTRW  & Continuous-time random walk \\
MFPT  & Mean first-passage time \\
MTA   & Mean time to absorption \\
mRNA  & Messenger ribonucleic acid \\
PTPP  & Persistent Telegraph Process with Pauses \\
WMFPT & Weighted mean first-passage time \\
\hline
\end{tabular}
\end{table}

\begin{longtable}{@{}p{0.38\textwidth}p{\dimexpr0.62\textwidth-2\tabcolsep\relax}@{}}
\caption{List of notation used throughout the text.}
\label{tab:notation}\\

\hline
\notationrow{\textbf{Symbol}}{\textbf{Meaning}}
\hline
\endfirsthead

\hline
\notationrow{\textbf{Symbol}}{\textbf{Meaning}}
\hline
\endhead

\hline
\endfoot
\hline

\multicolumn{2}{l}{\textbf{Persistent Telegraph Process with Pauses}} \\
\hline
\notationrow{$x$}{Position along a dendritic segment.}
\notationrow{$l$}{Length of a dendritic segment.}
\notationrow{$e_m\in\{+1,-1\}$}{Direction of motion; $+$ denotes anterograde and $-$ denotes retrograde motion.}
\notationrow{$r_{e_m}$, $p_{e_m}$}{Running state and pausing state associated with direction $e_m$.}
\notationrow{$s_m\in\{r_+,r_-,p_+,p_-\}$}{State of motion.}
\notationrow{$y=(x,s_m)$}{Deterministic state of the continuous-time Markov process.}
\notationrow{$Y_h=(X_h,S_{m,h})$}{Random state after a small time increment $h$.}
\notationrow{$f$}{Test function of the state.}
\notationrow{$\mathcal{T}_h$}{Markov semigroup operator.}
\notationrow{$\mathcal{L}$}{Infinitesimal generator.}
\notationrow{$v$}{Motor velocity during a running phase.}
\notationrow{$\lambda$}{Transition rate from the running state to the paused state.}
\notationrow{$\mu$}{Transition rate from the paused state to the running state.}
\notationrow{$\rho=\lambda/\mu$}{Ratio of running-to-pausing transition rates.}
\notationrow{$\alpha$}{Persistence probability after an anterograde run.}
\notationrow{$\beta$}{Persistence probability after a retrograde run.}
\notationrow{$\phi$}{Generic persistence probability; $\phi=\alpha$ for $e_m=+1$ and $\phi=\beta$ for $e_m=-1$.}
\notationrow{$T_a$}{First absorption time at either endpoint.}
\notationrow{$T_0$, $T_l$}{First hitting times of the left endpoint $0$ and right endpoint $l$.}

\hline
\multicolumn{2}{l}{\textbf{Single-segment absorption probabilities and first-passage times}} \\
\hline
\notationrow{$E\in\{0,l\}$}{Absorbing endpoint of a dendritic segment.}
\notationrow{$p^E_{s_m}(x)$}{Probability of absorption at endpoint $E$, starting from position $x$ in state $s_m$.}
\notationrow{$p^E_D(x)$, $p^E_S(x)$}{Difference and sum of absorption probabilities from the two running states.}
\notationrow{$w^E_{s_m}(x)$}{Weighted MFPT contribution to endpoint $E$, starting from $x$ in state $s_m$.}
\notationrow{$w^E_D(x)$, $w^E_S(x)$}{Difference and sum of the weighted MFPTs from the two running states.}
\notationrow{$c^E_{s_m}(x)$}{Conditional MFPT to endpoint $E$, given absorption at $E$.}
\notationrow{$A=(1-\alpha)\lambda/v$, \\ $B=(1-\beta)\lambda/v$,\\ $C=A/(Ae^{kl}-B)$, \\ $D=B/(Ae^{kl}-B)$,\\$ S=A+B$}{Auxiliary constants.}
\notationrow{$k=A-B$}{Exponential rate.}
\notationrow{$\eta=e^{kl}$}{Auxiliary parameter.}
\notationrow{$F_1,F_2,F_3,F_4$}{Forcing-term coefficients in the WMFPT equations.}
\notationrow{$C_1,C_2,\ldots,C_8$}{Integration constants in the WMFPT solutions.}
\notationrow{$\theta=1/(1+Al)$}{Auxiliary constant for the symmetric case $\alpha=\beta$.}

\hline
\multicolumn{2}{l}{\textbf{Semi-Markov chain on a network}} \\
\hline
\notationrow{$i,j,k,m$}{Indices of transient states or nodes.}
\notationrow{$s$}{Index of an absorbing state, interpreted as a synapse.}
\notationrow{$N$}{Total number of states in the semi-Markov chain.}
\notationrow{$p_{ij}$}{One-step transition probability from state $i$ to state $j$.}
\notationrow{$\mathbf{P}$}{Transition matrix of the embedded Markov chain.}
\notationrow{$\mathbf{Q}$}{Transient-to-transient submatrix of $\mathbf{P}$.}
\notationrow{$\mathbf{S}$}{Transient-to-absorbing submatrix of $\mathbf{P}$.}
\notationrow{$\mathbf{I}$}{Identity matrix.}
\notationrow{$\sigma_{is}$}{Path from transient state $i$ to absorbing state $s$.}
\notationrow{$\Sigma_{is}$}{Set of paths from $i$ to $s$.}
\notationrow{$P(\sigma_{is})$}{Probability of path $\sigma_{is}$.}
\notationrow{$L(\sigma_{is})$}{Length of a path, measured in the number of semi-Markov steps.}
\notationrow{$\mathcal{C}_{ij}$}{Random sojourn/traversal time associated with transition $i\to j$.}
\notationrow{$C_{ij}(\tau)$}{Density of the conditional sojourn time for transition $i\to j$.}
\notationrow{$c_{ij}=\mathbb{E}[\mathcal{C}_{ij}]$}{Mean sojourn/traversal time conditioned on transition $i\to j$.}
\notationrow{$r$}{Auxiliary variable.}
\notationrow{$a_{ij}=p_{ij}e^{r c_{ij}}$}{Element of the auxiliary matrix.}
\notationrow{$\mathbf{A}$}{Auxiliary matrix with entries $a_{ij}$.}
\notationrow{$\mathbf{A}_{\rm tt}$, $\mathbf{A}_{\rm ta}$}{Transient-to-transient and transient-to-absorbing blocks of $\mathbf{A}$.}
\notationrow{$\mathbf{B}_{\rm tt}$, $\mathbf{B}_{\rm ta}$}{Matrices with entries $p_{ij}c_{ij}$ for transient-to-transient and transient-to-absorbing transitions.}
\notationrow{$\mathbf{N}=(\mathbf{I}-\mathbf{Q})^{-1}$}{Fundamental matrix of the embedded Markov chain.}
\notationrow{$n_{ij}$}{Expected number of visits to state $j$ before absorption, for a walk starting at $i$ -- element of $\mathbf{N}$.}
\notationrow{$\bm{\Pi}=\mathbf{N}\mathbf{S}$}{Matrix of absorption probabilities.}
\notationrow{$\pi_{is}$}{Probability of absorption at synapse $s$, starting from $i$.}
\notationrow{$p_{ii}^{(j)}$}{Probability of returning to the starting node $i$ after entering segment $i\to j$.}
\notationrow{$\bm{\omega}_s$}{WMFPT vector for absorption at $s$.}
\notationrow{$\bm{\Omega}$}{WMFPT matrix.}
\notationrow{$\omega_{is}$}{WMFPT from state $i$ to synapse $s$.}
\notationrow{$\gamma_{is}=\omega_{is}/\pi_{is}$}{CMFPT from state $i$ to synapse $s$.}
\notationrow{$d_{ks}$}{Target-weighted mean sojourn-time contribution at transient state $k$ associated with eventual absorption at $s$.}
\notationrow{$d_k$}{Mean waiting time at state $k$.}
\notationrow{$\mathbf{D}$}{Matrix with elements $d_{ks}$.}
\notationrow{$\mathbf{u}$}{Column vector of ones.}
\notationrow{$\Lambda_1$}{Dominant eigenvalue of $\mathbf{Q}$.}
\notationrow{$\Lambda_2$}{Second-largest eigenvalue of $\mathbf{Q}$ in modulus.}

\hline
\end{longtable}

\section{Persistent Telegraph Process with Pauses}
\label{sec:ptpp}

\subsection{Setup and absorption probabilities}
\label{sec:ptpp_absorption}

Consider a Persistent Telegraph Process with Pauses describing the random motion of molecular cargo on the line segment $[0, l]$. Let $\alpha$ and $\beta$ denote the persistence probabilities for the anterograde and retrograde directions, respectively. Assume that at a given moment, which for convenience can be taken to be $t=0$, the cargo is at position $x$ in the running state $r_{e_m}$, corresponding to the state $(x, r_{e_m})$ in the continuous-time Markov chain. For convenience, the notation $e_m$ will be used for a sign, $+$ or $-$, or a signed number, $+1$ or $-1$. After a very small time interval $h$, the cargo continues moving in the same direction with probability $1-\lambda h$, and stops with probability $\lambda h$\footnote{While $dt$ is the natural choice for an infinitesimal time increment, the notation $h$ is used as in the mathematical literature on the topic.}. The respective future state in the Markov process could thus be a running state $(X_h, r_{e_m})$ with probability $1-\lambda h$, or a pausing one $(x, p_{e_m})$ with probability $\lambda h$, where $p_{e_m}$ denotes a pause after a run in direction $e_m$. We have disregarded two or more transitions since the respective probability in the interval $(0,h)$ is of order $o(h)$. If motion continues, the future position of the cargo is
\begin{equation} 
X_h = x + e_m v h.
\end{equation}
Because motion is assumed to have constant velocity during running phases, randomness arises only from state transitions. Now, consider a test function of the state, $f$, for which the expansion of the expectation at the future state is
\begin{equation}
\mathbb{E}_{(x,r_{e_m})}[f(X_h,S_{m,h})] \approx (1-\lambda h)[f(x, r_{e_m}) + \frac{\partial f}{\partial x} (X_h-x)] + \lambda h f(x, p_{e_m}),
\end{equation} 
where $S_{m,h}$ is the random future state of motion, which can be either $r_{e_m}$ or $p_{e_m}$. The expansion retains terms through first order in $h$; contributions from two or more state transitions are of higher order. Now, by discarding the term containing $h^2$, one has 
\begin{equation}
\mathbb{E}_{(x,r_{e_m})}[f(X_h,S_{m,h})] - f(x, r_{e_m}) = e_mvh \frac{\partial f}{\partial x} + \lambda h [f(x, p_{e_m}) - f(x,r_{e_m})] + o(h).
\end{equation}
The generator for a running state $r_{e_m}$ is then
\begin{equation}
(\mathcal{L}^{(r_{e_m})} f)(x, r_{e_m}) = e_mv \frac{\partial f}{\partial x} + \lambda [f(x, p_{e_m}) - f(x,r_{e_m})].\label{sup:eq:gen_move}
\end{equation}
One can use the same approach for the paused state and take into account that the cargo can start running in the same direction as the previous one with a given persistence probability $\phi$, or in the opposite direction with probability $1-\phi$. Since the transition rate parameter of the pause is $\mu$, the expectation of the test function at the future state is given by
\begin{equation}
\mathbb{E}_{(x,p_{e_m})}[f(X_h, S_{m,h})] = (1-\mu h) f(x, p_{e_m}) +\mu h\left[\phi f(x, r_{e_m}) + (1 - \phi) f(x,r_{-e_m})\right] + o(h),
\end{equation}
where $-e_m$ is the direction opposite to $e_m$. Using this relationship in the definition of the generator gives
\begin{equation}
(\mathcal{L}^{(p_{e_m})} f)(x, p_{e_m}) = \mu\left[\phi f(x, r_{e_m}) + (1 - \phi) f(x,r_{-e_m})-f(x, p_{e_m})\right],\label{sup:eq:gen_rest}
\end{equation}
where the probability is $\phi = \alpha$ for the anterograde direction, while $\phi = \beta$ for the retrograde one. 

To obtain the generator equations for the absorption probabilities and WMFPTs, one should use the respective test functions $f=p$ and $f=w$ and the specific states of motion in \eqref{sup:eq:gen_move} and \eqref{sup:eq:gen_rest}. Since probability is conserved, the action of the generator on the absorption probability test function is zero, while for the WMFPTs there is a nonzero forcing term, which is given in the main text.

\subsection{Solution of the generator equations for the WMFPT}
\label{sec:ptpp_wmfpt_solution}

Consider the case of the right endpoint $l$ and use the absorption probabilities in the equations for the sum and difference \eqref{main:eq:SD_WMFPT_ODE}. The differential equation for the difference of WMFPTs takes the form
\begin{equation}
\frac{dw^l_D(x)}{dx} = k w^l_D(x) - F_1 e^{kx} +F_2,
\end{equation}
where the forcing terms are
\begin{eqnarray}
F_1 = \frac{1}{v}\left[\rho(\alpha-\beta)(C-D) + (1+\rho)(C+D)\right]; \hspace{1cm} F_2 = \frac{2D(1+\rho)}{v}.
\end{eqnarray}
The solution of the last differential equation is 
\begin{equation}
w^l_D(x) = (C_1 - F_1x)e^{kx} - \frac{F_2}{k}.
\end{equation}
The last result can be used in the equation for the sum
\begin{equation}
\frac{dw^l_S(x)}{dx} = (A+B) w^l_D(x) + F_3 e^{kx},
\end{equation}
where the forcing function parameter is
\begin{equation}
F_3 = \frac{1}{v}\left[\rho(1-\alpha-\beta) - 1\right](C-D).
\end{equation}
The sum of WMFPTs is thus
\begin{equation}
w^l_S(x) = \frac{(A+B)(kC_1 - kF_1x + F_1)+kF_3}{k^2} e^{kx} - \frac{(A+B)F_2}{k}x+C_2.
\end{equation}
The integration constants are obtained from the boundary conditions $w^l_{r-}(0) = 0$ and $w^l_{r+}(l) = 0$. The respective values are
\begin{eqnarray}
C_1 &=& \frac{F_1\left[S(1-E)+2AlkE\right]+F_2k(Sl+2)+F_3k(1-E)}{2k(AE-B)} \nonumber\\
C_2 &=& -\frac{2BC_1}{k} - \frac{SF_1}{k^2}-\frac{F_2+F_3}{k},
\end{eqnarray}
where $S=A+B$ and $E = e^{kl}$.

The other endpoint is handled in the same way. 
The equation for the difference is
\begin{equation}
\frac{dw^0_D(x)}{dx} = k w^0_D(x) + F_1 e^{kx} -F_4,
\end{equation}
where $F_1$ is the same as for the other endpoint, while
\begin{equation}
F_4 = \frac{2C(1+\rho)e^{kl}}{v}.
\end{equation}
Its solution has a similar form 
\begin{equation}
w^0_D(x) = (C_3 + F_1x)e^{kx} + \frac{F_4}{k}.
\end{equation}
The equation for the sum is
\begin{equation}
\frac{dw^0_S(x)}{dx} = S w^0_D(x) - F_3 e^{kx} ,
\end{equation}
with solution
\begin{equation}
w^0_S(x) = \frac{SkC_3 - kF_3 + SF_1kx - SF_1}{k^2} e^{kx} + \frac{SF_4}{k}x+C_4.
\end{equation}
The boundary conditions are $w^0_{r+}(l)=w^0_{r-}(0)=0$. By applying the same substitutions $S=A+B$ and $E=e^{kl}$, the integration constants are
\begin{eqnarray}
C_3 &=& \frac{(E-1)(kF_3 + SF_1)-2AklF_1E-2F_4k-F_4Slk}{2k(AE-B)}; \nonumber \\
C_4 &=& \frac{k(F_3 + F_4) + SF_1 - 2kBC_3}{k^2}.\nonumber\\
\end{eqnarray}
Now, we have all the information needed to calculate the WMFPTs to each endpoint and use them to find the CMFPTs.

\subsection{Long-segment asymptotics}
\label{sec:ptpp_asymptotics}

For \(\alpha\neq\beta\), the long-segment asymptotic regime is characterized by $|k|l\gg1$, 
where \(k=A-B=(\lambda/v)(\beta-\alpha)\). Since \(v/\lambda\) is the typical run length, this condition is stronger than \(l\gg v/\lambda\) when \(\alpha\) and \(\beta\) are close. Then, for $\alpha \neq \beta$ one has either $e^{kl} \to 0$ or $e^{kl} \to \infty$. Table~\ref{tab:asymptotic_parameters} summarizes the asymptotic behavior as $l\to \infty$ for both cases.
\begin{table}[ht]
\caption{Asymptotic behavior of the parameters involved in calculating the CMFPT for traversal of a dendritic segment.}
\label{tab:asymptotic_parameters}
\centering
\begingroup
\renewcommand{\arraystretch}{1.25}
\begin{minipage}[t]{0.42\textwidth}
\centering
\begin{tabular}{c|c|c}
\hline
Quantity & $k>0$ & $k<0$ \\
\hline
$A$ & $O(1)$ & $O(1)$ \\
$B$ & $O(1)$ & $O(1)$ \\
$k$ & $O(1)$ & $O(1)$ \\
$S=A+B$ & $O(1)$ & $O(1)$ \\
\hline
$\eta=e^{kl}$ & $\to \infty$ & $\to 0$ \\
\hline
$C=\dfrac{A}{A\eta-B}$
& $\sim e^{-kl}$
& $\sim -\dfrac{A}{B}$ \\
\hline
$D=\dfrac{B}{A\eta-B}$
& $\sim \dfrac{B}{A}e^{-kl}$
& $\sim -1$ \\
\hline
\end{tabular}
\end{minipage}
\hfill
\begin{minipage}[t]{0.54\textwidth}
\centering
\begin{tabular}{c|c|c}
\hline
Quantity & $k>0$ & $k<0$ \\
\hline
$F_1$ & $O(e^{-kl})$ & $O(1)$ \\
$F_2$ & $O(e^{-kl})$ & $O(1)$ \\
$F_3$ & $O(e^{-kl})$ & $O(1)$ \\
\hline
$C_1$ & $O(le^{-kl})$ & $O(l)$ \\
$C_2$ & $O(le^{-kl})$ & $O(l)$ \\
\hline
\end{tabular}
\end{minipage}
\endgroup
\end{table}

Using the asymptotic values for favored retrograde motion, $\beta > \alpha$, or $k>0$, gives $w^l_{r+}(0) \approx C_1$, which, combined with $p^l_{r+}(0) = C-D$, results in
\begin{equation}
c^l_{\beta>\alpha} \approx \frac{(1+\rho)(2-\alpha-\beta)-\rho(\beta-\alpha)^2}{v(\beta-\alpha)}l.
\end{equation}
For the opposite case, $k<0$, the asymptotic behavior is
\begin{equation}
c^l_{\alpha>\beta} \approx \frac{(1+\rho)(2-\alpha-\beta)}{v(\alpha-\beta)}l.
\end{equation}
    
\subsection{Symmetric persistence}
\label{sec:ptpp_symmetric}

Consider the case in which $\alpha = \beta$, which further implies $A=B$. Substituting this into the matrix equation \eqref{main:eq:Abs_p_matrix} and then in the relationships between absorption probabilities \eqref{main:eq:abs_prob_rest_from_m} gives
\begin{equation}
\begin{aligned}
p_{r+}^l(x) &= \frac{1 + A x}{1 + A l}; 
&\qquad
p_{r-}^l(x) &= \frac{A x}{1 + A l}; \\
p_{p+}^l(x) &= \frac{\alpha + A x}{1 + A l}; 
&\qquad
p_{p-}^l(x) &= \frac{1 - \alpha + A x}{1 + A l}; \\
p_{r+}^0(x) &= \frac{A(l - x)}{1 + A l}; 
&\qquad
p_{r-}^0(x) &= \frac{1 + A(l - x)}{1 + A l}; \\
p_{p+}^0(x) &= \frac{1 - \alpha + A(l - x)}{1 + A l}; 
&\qquad
p_{p-}^0(x) &= \frac{\alpha + A(l - x)}{1 + A l}.
\end{aligned}  \label{sup:eq:Abs_prob_sym}
\end{equation}
The system of equations for the sum and difference of WMFPTs \eqref{main:eq:SD_WMFPT_ODE} simplifies to
\begin{eqnarray}
\frac{dw^E_D(x)}{dx} &=&  -\frac{1+\rho}{v}p^E_S(x); \nonumber\\
\frac{dw^E_S(x)}{dx} &=& 2Aw^E_D(x) +\frac{1}{v}\left[\rho(1-2\alpha)- 1\right] p^E_D(x) . \label{sup:eq:SD_WMFPT_ODE_sym}
\end{eqnarray}

Before solving this system, note that the sums and differences of absorption probabilities for both endpoints are
\begin{equation}
\begin{aligned}
p_S^l(x) &= \theta(1 + 2Ax), 
&\qquad
p_D^l(x) &= \theta, \\
p_S^0(x) &= \theta[1 + 2A(l - x)], 
&\qquad
p_D^0(x) &= -\theta,
\end{aligned}
\label{sup:eq:Abs_prob_SD_sym}
\end{equation}
where $\theta = 1/(1+Al)$. Using the first pair in \eqref{sup:eq:SD_WMFPT_ODE_sym} gives the general solution for the right endpoint
\begin{align}
w_D^l(x) &= C_5 - \frac{(1+\rho)\theta}{v}\left(x + A x^2\right), \nonumber  \\[6pt]
w_S^l(x) &= C_6 + 2A C_5 x 
- \frac{A(1+\rho)\theta}{v} x^2 
- \frac{2A^2(1+\rho)\theta}{3v} x^3 
+ \frac{\theta}{v}\left[\rho(1 - 2\alpha) - 1\right] x.
\end{align}
The boundary conditions $w^l_{r+}(l)=0$ and $w^l_{r-}(0)=0$ are satisfied by the constants
\begin{equation}
C_5=C_6 = \frac{\theta l}{v(1 + Al)}
\left[1 + \alpha \rho + Al(1 + \rho) + \frac{A^2 l^2}{3}(1 + \rho) \right].
\end{equation}
Likewise, for the left endpoint one has the difference and sum
\begin{eqnarray}
w_D^0(x) &=& C_7 -\frac{(1+\rho)\theta}{v}\left[(1+2Al)x-Ax^2\right], \\
w_S^0(x) &=& C_8 + 2A C_7 x
-\frac{A(1+\rho)\theta}{v}(1+2Al)x^2
+\frac{2A^2(1+\rho)\theta}{3v}x^3
-\frac{\theta}{v}\bigl[\rho(1-2\alpha)-1\bigr]x. \nonumber
\end{eqnarray}
The integration constants coincide again and follow from the boundary conditions $w^0_{r+}(l)=0$ and $w^0_{r-}(0)=0$
\begin{equation}
C_7=C_8 = \frac{\theta^2l}{v}
\left[\rho(1-\alpha) + Al(1+\rho) +\frac{2}{3}A^2l^2(1+\rho) \right].
\end{equation}

\section{Derivation of the matrix-form WMFPT equation in the semi-Markov model}
\label{sec:semi_markov}

Differentiating the matrix product in \eqref{main:eq:MFPT_all} results in
\begin{equation}
\bm{\Omega} = \left[\frac{\partial \left(\mathbf{I}-\mathbf{A}_{\rm tt}\right)^{-1}}{\partial r}\mathbf{A}_{\rm ta}  + \left(\mathbf{I}-\mathbf{A}_{\rm tt}\right)^{-1}\frac{\partial \mathbf{A}_{\rm ta}}{\partial r}\right]\Big|_{r=0}.
\end{equation}
Using the derivative of a matrix inverse gives
\begin{equation}
\bm{\Omega} =\left[\left(\mathbf{I}-\mathbf{A}_{\rm tt}\right)^{-1}\frac{\partial \mathbf{A}_{\rm tt}}{\partial r}\left(\mathbf{I}-\mathbf{A}_{\rm tt}\right)^{-1}\mathbf{A}_{\rm ta}\right]\Big|_{r=0} + \left[\left(\mathbf{I}-\mathbf{A}_{\rm tt}\right)^{-1}\frac{\partial \mathbf{A}_{\rm ta}}{\partial r}\right]\Big|_{r=0}
\end{equation}
This expression can be simplified further by observing that the calculation of the inverse involves products of elements $a_{ik}$, which, at $r=0$, become the corresponding products of transition probabilities $p_{ik}$. Thus, by using the notation in the canonical form \eqref{eq:CAN_Form} one has
\begin{equation}
\left[\left(\mathbf{I}-\mathbf{A}_{\rm tt}\right)^{-1}\right]\Big|_{r=0} = \left(\mathbf{I}-\mathbf{Q}\right)^{-1}; \hspace{0.5 cm} \mathbf{A}_{\rm ta}\Big|_{r=0} = \mathbf{S}.
\end{equation}
The derivatives of the elements of the matrix $\mathbf{A}_{\rm tt}$ at $r=0$ are $b_{ik} = p_{ik}c_{ik}$, which can be organized in a matrix $\mathbf{B}_{\rm tt}$. The same procedure can be applied to the derivative of the matrix $\mathbf{A}_{\rm ta}$ to obtain the matrix $\mathbf{B}_{\rm ta}$. To simplify the notation, we note that the inverse 
\begin{equation}
\left(\mathbf{I}-\mathbf{Q}\right)^{-1} = \mathbf{N} \label{sup:eq:N_fund}
\end{equation}
is known as the fundamental matrix for the classical absorbing Markov chain \cite{grinstead2012introduction}, whose entries $n_{ik}$ represent the mean number of visits to state $k$ for a walk that started at $i$. Now the WMFPT matrix is
\begin{equation}
\bm{\Omega} = \mathbf{N}\mathbf{B}_{\rm tt} \mathbf{N}\mathbf{S}+
\mathbf{N}\mathbf{B}_{\rm ta}.
\end{equation}
The element $\pi_{is}$ of the product $\Pi = \mathbf{N}\mathbf{S}$ represents the probability of absorption at target $s$ by a walk starting at transient node $i$ \cite{grinstead2012introduction}
\begin{equation}
\pi_{is} = \sum_{k \in \mathcal{T}} n_{ik} p_{ks}.
\end{equation}
Since each walk is eventually absorbed, one has the normalization $\sum_s \pi_{is} = 1$. Thus, we have a slightly simpler relationship for the WMFPTs
\begin{equation}
\bm{\Omega} = \mathbf{N}\left(\mathbf{B}_{\rm tt} \Pi+\mathbf{B}_{\rm ta}\right).
\end{equation}
The element of the matrix expression $\mathbf{B}_{\rm tt} \Pi+\mathbf{B}_{\rm ta}$ is
\begin{equation}
d_{ks} =  \sum_{m \in \mathcal{T}} p_{km} c_{km} \pi_{ms} + p_{ks} c_{ks}. \label{sup:eq:MWT_matr_elem}
\end{equation}
Summing over all targets $s$ gives
\begin{equation}
d_{k} = \sum_{m \in \mathcal{T}} p_{km} c_{km} \sum_s \pi_{ms}  + \sum_s p_{ks} c_{ks} = \sum_{m \in \mathcal{T}} p_{km} c_{km} + \sum_s p_{ks} c_{ks},
\end{equation}
where we have applied the normalization $\sum_s \pi_{ms}=1$. It is obvious that $d_k$ represents the sojourn time at transient node $k$, accounting for jumps to all neighbors, whether transient or absorbing. Thus, $d_{ks}$ represents the target-weighted mean sojourn-time contribution generated at transient state $k$ and associated with eventual absorption at $s$. Note that $d_{ks}$ is weighted because the waiting time is multiplied by the probability of occurrence, with trajectories constrained to be absorbed at the target $s$. So, one can write a simple matrix equation for the WMFPTs
\begin{equation}
\bm{\Omega} = \mathbf{ND}, \label{sup:eq:WMFPT_matrix}
\end{equation}
where $\mathbf{D}$ is the waiting-time matrix with elements $d_{ks}$. The relationship \eqref{sup:eq:WMFPT_matrix} reduces to the classical result with constant waiting time. By taking $c_{ik}=1$, one has
\begin{equation}
\bm{\Omega} = \mathbf{N}\mathbf{Q} \mathbf{N}\mathbf{S}+
\mathbf{N}\mathbf{S} = \mathbf{N}(\mathbf{Q} \mathbf{N} + \mathbf{I})\mathbf{S}.
\end{equation}
Notice that $\mathbf{Q} \mathbf{N} + \mathbf{I} = \mathbf{N}$, which can be verified by multiplying both sides on the right by $\mathbf{N}^{-1}$ and using \eqref{sup:eq:N_fund}. Thus, the WMFPTs to different absorbing states with a fixed waiting time satisfy the simple relationship
\begin{equation}
\bm{\Omega} = \mathbf{N}^2\mathbf{S}=\mathbf{N}\Pi.\label{sup:eq:MFPT_class_multiple}
\end{equation}
In the simplest case of a search for one target, the column vector is the vector of ones, $\Pi = \mathbf{u}$, since the probability of absorption at the only target is one. This recovers the familiar result
\begin{equation}
\bm{\Omega} = \mathbf{Nu}.\label{sup:eq:WMFPT_class_single}
\end{equation}
Finally, the CMFPT, which is the biologically relevant quantity because it represents the expected age of an mRNA molecule, is obtained from \eqref{eq:MFPTs_relationship}.

\section{Numerical implementation}
\label{sec:numerics}

The morphological data of each neuron provide enough information to reconstruct its shape and dimensions. In this work, the simplest representation was implemented, so only dendritic segment lengths were considered, without accounting for spatial orientation. Thus, the dendritic tree was constructed with the soma, branching points, and synapses as nodes and dendritic segments as edges. Because explicit synaptic locations are not provided in the neuronal reconstructions used here, terminal dendritic tips are designated as the absorbing synaptic targets. The theoretical values of the CMFPT to synapses are calculated for the dendritic lengths obtained from the files.

For the Monte Carlo simulations examining the CMFPT and traversal probabilities along a hypothetical dendritic segment, each simulation run was initiated at position $x=10^{-15}\,\mu\mathrm{m}$, moving in the positive direction. The results presented in the main text are obtained as averages of 1 000 000 simulation runs.

\begin{figure}[t]
\centering

\begin{minipage}{0.48\textwidth}
    \centering
    \textbf{(a)}\\[-0.3em]
    \includegraphics[width=\linewidth]
    {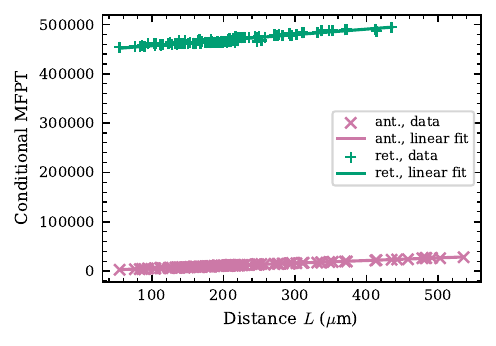}
\end{minipage}
\hfill
\begin{minipage}{0.48\textwidth}
    \centering
    \textbf{(b)}\\[-0.3em]
    \includegraphics[width=\linewidth]
    {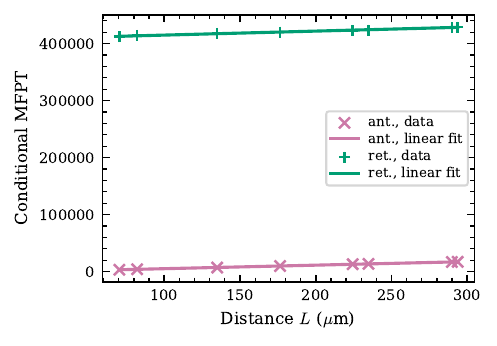}
\end{minipage}

\vspace{0.45cm}

\begin{minipage}{0.48\textwidth}
    \centering
    \textbf{(c)}\\[-0.3em]
    \includegraphics[width=\linewidth]
    {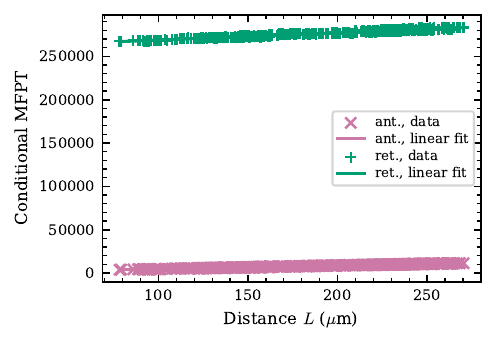}
\end{minipage}
\hfill
\begin{minipage}{0.48\textwidth}
    \centering
    \textbf{(d)}\\[-0.3em]
    \includegraphics[width=\linewidth]
    {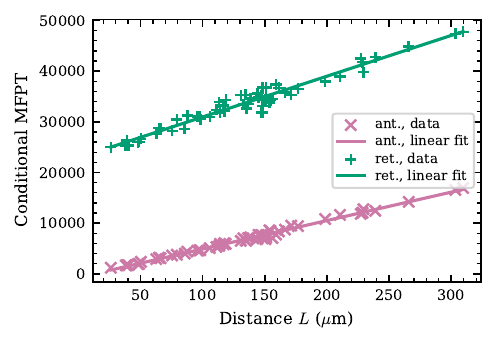}
\end{minipage}

\vspace{0.45cm}

\begin{minipage}{0.48\textwidth}
    \centering
    \textbf{(e)}\\[-0.3em]
    \includegraphics[width=\linewidth]
    {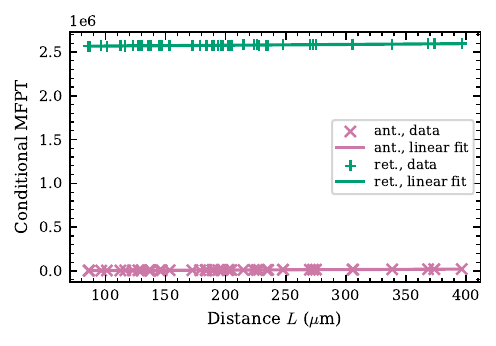}
\end{minipage}
\hfill
\begin{minipage}{0.48\textwidth}
    \centering
    \textbf{(f)}\\[-0.3em]
    \includegraphics[width=\linewidth]
    {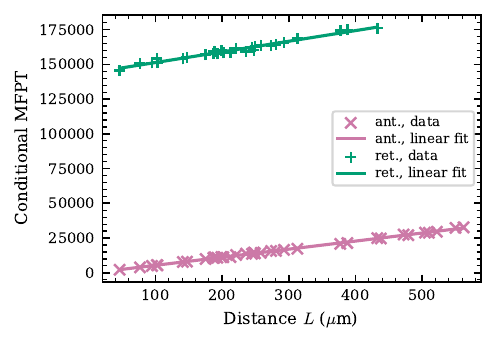}
\end{minipage}

\caption{
Conditional mean first-passage time (CMFPT) as a function of the
soma-to-synapse path distance for six neuronal reconstructions.
The panels correspond to:
(a) dorsal hippocampal CA1 pyramidal cell of a rat (NMO\_83166) \cite{dalrymple2015anterior};
(b) dentate gyrus granule cell of a mouse (NMO\_06353) \cite{murphy2011heterogeneous};
(c) cerebellar Purkinje cell of a mouse (NMO\_35058) \cite{anwar2014dendritic}; 
(d) CA1 pyramidal neuron of a rat with postnatal fluoxetine treatment (NMO\_158968) \cite{mukhopadhyay2021postnatal};
(e) dorsal CA1 pyramidal neuron of a rat, right hippocampus (NMO\_158944)    \cite{mukhopadhyay2021postnatal};
(f) dorsal CA1 pyramidal neuron of a rat, left hippocampus (NMO\_158955) \cite{mukhopadhyay2021postnatal}.
Solid lines show linear fits of the calculated CMFPT to the synapses.}
\label{fig:supplementary_CMFPT_neurons}
\end{figure}

To verify the general observation of a linear dependence of the CMFPT on distance, Fig.~\ref{fig:supplementary_CMFPT_neurons} shows the calculations for six neuron morphologies. Details of the neurons are provided in the following table.

\begin{table}[t]
\caption{Neuronal reconstructions obtained from NeuroMorpho.Org
and used in the CMFPT calculations.}
\label{tab:neuromorpho_cells}
\centering
\begin{tabularx}{\textwidth}{@{}>{\raggedright\arraybackslash}X >{\raggedright\arraybackslash}p{0.22\textwidth} >{\raggedright\arraybackslash}p{0.25\textwidth}@{}}
Cell name & NeuroMorpho.Org ID & Archive \\
\hline
0001-11\_1
    & NMO\_83166 & Harland \cite{dalrymple2015anterior} \\

DGC671
    & NMO\_06353 & Danzer \cite{murphy2011heterogeneous} \\

10-2012-02-09-001
    & NMO\_35058 & De Schutter \cite{anwar2014dendritic} \\ 

PNFlx-Slide-1-Section-2-L-CA1-pyr-neuron-1
    & NMO\_158968 & Vaidya \cite{mukhopadhyay2021postnatal} \\

Vehicle-Slide-1-Section-2-R-dCA1-pyr-neuron-1
    & NMO\_158944 & Vaidya \\

Vehicle-Slide-3-Section-3-L-dCA1-pyr-neuron-1
    & NMO\_158955 & Vaidya
\end{tabularx}
\end{table}
The figure confirms the observed pattern: the constant term of the linear fit for stronger retrograde persistence is substantial and, in certain cases, dwarfs the differences in the CMFPT. The theory of nearly equal travel times to synapses is thus applicable to the explanation of this phenomenon.

